\documentclass[12pt]{article}

\usepackage[textwidth=6.5in,textheight=9in]{geometry}
\usepackage{amsmath,graphicx,tipa,amssymb}

\usepackage{authblk}

\newcommand{\PP}{ \mathbb{P} }
\newcommand{\EE}{ \mathbb{E} }
\newcommand{\VV}{ \mathbb{V} }

\newcommand{\rar}{\rightarrow}
\newcommand{\bv}[1]{\boldsymbol{\mathbf{#1}}}

\newcommand{\la}{\langle}
\newcommand{\ra}{\rangle}

\title{\textbf{Inferring the drivers of language change using spatial models}  }

\author[1]{James Burridge}
\author[2]{Tamsin Blaxter}

\affil[1]{School of Mathematics and Physics, University of Portsmouth, Portsmouth PO1 3HF, United Kingdom}
\affil[2]{Gonville and Caius College, Cambridge CB2 1TA, United Kingdom}

\begin{document}

\maketitle

\begin{abstract}
Discovering and quantifying the drivers of language change is a major challenge. Hypotheses about causal factors proliferate, but are difficult to rigorously test. Here we ask a simple question: can 20th Century changes in English English be explained as a consequence of spatial diffusion, or have other processes created bias in favour of certain linguistic forms? Using two of the most comprehensive spatial datasets available, which measure the state of English at the beginning and end of the 20th century, we calibrate a simple spatial model so that, initialised with the early state, it evolves into the later. Our calibrations reveal that while some changes can be explained by diffusion alone, others are clearly the result of substantial asymmetries between variants. We discuss the origins of these asymmetries and, as a by-product, we generate a full spatio-temporal prediction for the spatial evolution of English features over the 20th Century, and a prediction of the future.   
\end{abstract}

%
%
%
%
%

\section{Modelling language evolution}

Modelling the collective behaviour of systems of sentient agents, from flocks of birds \cite{cas09,bia12}, to economies \cite{yak00}, cities \cite{bar19} and languages \cite{bax06,bur16,bur17}, is attractive, but not easy. To quote Emanuel Derman \cite{der07}, an early adopter of financial modelling,  

\vspace{0.25cm}
\textit{In physics there may one day be a Theory of Everything; in finance and the social sciences, you’re lucky if there is a usable theory of anything. }
\vspace{0.25cm}

\noindent
The problems which arise when trying to model social systems depend to some extent on the system in question, but share a good deal in common. In many cases we have only one realisation of the change process which we want to understand, and the data available about the change may be sparse. Because we cannot observe the thought processes and motivations of the agents, we caricature them with a simple model. However, many different models may fit observations similarly well. This problem recedes if we have more data and more realisations, which allow us to apply Statistical Learning methods \cite{has09}, such as cross validation, to rigorously select the ``best model'' by maximizing its ability to predict unseen data. If this is not possible, because we have only one realisation of a particular change process, then another approach is to start with some simple default theory, or \textit{null model}  \cite{bly12_2}, and then seek to determine whether it is sufficient to explain observations. We may also consider an \textit{alternative model} with incrementally increased complexity and minimal additional assumptions, the aim being to understand what is missing from the null model, while avoiding unjustified assumptions about what should replace it.

The broad question which interests us is: what are the processes which drive the evolution of languages, and how should they be represented in a coarse grained spatial model? A longer term aim is to be in possession of sets stochastic equations, derived from simple assumptions about speaker-level behaviour, which describe how language evolves over space and time, and amongst different social groups. Such equations should calibrate to available data on the language state up until the present, then allow us to understand historic changes, and make future predictions. Here we take a step toward addressing these aims, by fitting a coarse grained spatial model to two large scale spatial datasets which capture the states of various English language features at the beginning and end of the 20th Century.

The mathematical modelling of language evolution, often using methods of statistical physics, is now an established field. Many aspects of language change have been studied using non-spatial models \cite{now01,bax06,mit11,mit17,bly07}, and an important focus has been to understand the replacement of one linguistic feature by another, typically following the ubiquitous temporal S-curve \cite{bly12,gha14}. Such changes can be driven by numerous mechanisms, including those which favour particular variants (e.g. regularization \cite{lie07}, social biases \cite{cha98}, or exogenous factors \cite{gha14}), by stochastic effects \cite{kar20,new17}, and by biases driven by linguistic variation over the age spectrum  \cite{mit11,mit17,sta16,lab01,bax16}.  In this work we allow speakers a learning bias toward individual linguistic features, without explicitly modelling any particular mechanisms which might generate this. We seek only to determine whether some form of bias existed, given the observed changes.

Spatial modelling of language change is a topic of growing interest \cite{kau21,sta13,ner10, tes06, bar06_2, bur17, bur18,bur19,bur20}. Whereas non-spatial linguistic data is plentiful, detailed  spatial datasets are rarer. Their construction typically  requires either large scale collaborative effort \cite{ort62,ellis1889,hee01}, or viral success online \cite{lee16, leemann2017, vau17}. Spatial modelling presents additional challenges in the form of greater analytical and computational complexity, particularly when matching models to real world spatial language distributions. There is also long standing interest within the quantitative linguistics community in the statistical analysis of spatial, social and temporal variations within real linguistic domains \cite{hee01,gri11,gri16}.

Recent spatial models \cite{bur17,bur18,bur19,bur20}, which accurately model geography and population distributions, suggest that the shapes of linguistic domains and the locations of cities and towns can have a partially predictable influence on the evolution of language. In fact, the importance of geography has been well known to dialectologists for a long time \cite{cha98, blo33}. From a mathematical perspective, this influence arises from the impact of geography on the shapes of boundaries (isoglosses) between alternative linguistic features, which behave like two dimensional bubbles. The existence of this surface tension effect has recently been tested using historical English dialect data \cite{bur20}.  We therefore believe that space and spatial processes must be part of models which seek to understand language change processes. Here we provide, to our knowledge, the first detailed spatial model which matches the evolution of real linguistic features within an extended spatial domain (England in the 20th Century), accounting for realistic migration patterns, and plausible learning behaviour.

\section{Data and existing theories}

\subsection{Linguistic variables and variants}

For the purpose of academic study, languages may be broken down into distinct components:  single units of sounds (\textit{phones}), rules for combining sounds  (\textit{phonology}), words (the \textit{lexicon}), rules for word construction (\textit{morphology}), and rules for sentence construction (\textit{syntax}). A language may be viewed as a complete specification of all its components, and language change as the process of progressively exchanging components for alternatives. We refer to a language feature for which there are a set of alternatives as a \textit{variable}, and to the alternatives as \textit{variants}. These might be different words for the same object or idea, different sounds playing the same role in certain groups of words, or alternative syntactic rules. As an example, the word for the season after summer historically had variants \textit{autumn}, \textit{backend} and \textit{fall} in England \cite{ort62}. The fundamental quantities which we build models of are the relative usage frequencies of variants at different locations (or regions) in space. Simultaneous spatial variations in many linguistic variables can create distinctive regional \textit{dialects}. 

\subsection{The Survey of English Dialects and the English Dialect App}

We model the spatial evolution of a set of linguistic variables over the 20th Century in England, the largest and most densely populated nation within the British mainland. As initial and final conditions, we use data for the same variables from two surveys: the Survey of English Dialects (SED) \cite{orton1962} and the English Dialects App (EDA) \cite{lee16,leemann2017}. 

The SED was carried out in the 1950s at 313 localities. With 986 respondents, it is a small survey by modern standards. The localities were selected to provide a relatively evenly-spaced sample across England, but included almost no speakers from urban areas. This was a deliberate choice; the survey sought out the most conservative demographic (typically farm-labourers born in the 1870s and 80s) to capture the network of ``traditional'' dialects before they disappeared. Methodologically, the SED was rigorous, making few allowances for speed or convenience at the expense of data volume and quality: fieldworkers visited respondents in person and recorded their responses to over a thousand questions in narrow phonetic transcription (and later also on tape). Despite some criticisms arising from systematic differences in how fieldworkers asked questions or transcribed responses \cite{mathussek2016,payne2017}, the SED is a rich and relatively trustworthy source of data and has provided the material for more than half a century's worth of secondary analysis.

The EDA represents a new generation of dialect data collection methods, using digital technologies to reach a large number of respondents very cheaply. Over 50,000 speakers answered 26 questions about their usage through a smartphone app in 2016, and all but one question duplicated variables surveyed in the SED. The EDA exemplifies a different set of trade-offs between data quality, quantity and speed. It is possible to mine the SED data for patterns across many responses and examine variables which were not specifically targeted, whereas the EDA required speakers to decide between predefined categories, limiting analysis to 26 variables and predetermined sets of variants. Instead, the EDA's major advantage is its 45,287 speakers in 39,590 locations within the region covered by the SED. It is also more representative, covering urban areas and all demographics, although, because it was carried out via smartphones, it is somewhat skewed towards younger and more affluent speakers.

In spite of methodological differences, we argue that comparison of the SED and EDA is a viable way to explore language change in 20th century English English. Laboratory perception tests \cite{leemann2017}  suggest that speakers are largely able to discriminate between the variants of the phonetic variables elicited in the EDA. Spatial patterns in the results of the EDA, when compared to the SED, show that levelling (loss of variants) and isogloss (linguistic boundary) movements are in line with our prior understanding of the mechanisms of language change. However, because typical SED respondents were old, while EDA respondent were younger, we are not simply mapping the 60 or so years of language change between the median dates of the two language surveys.  If we assume that speakers' linguistic norms are set early in life and change little in adulthood (the underlying assumption of much historical linguistic \cite{ringe2013} and sociolinguistic \cite{labov1994} work on language change), the relevant comparison is instead speakers' dates of birth: the median date of birth is the 3rd of May 1881 for SED respondents, and the 20th of July 1983 for EDA respondents, implying that we are looking at around 100 years of change. 

Because urban areas are under-represented in the SED, we must consider whether city varieties are likely to have differed substantially from the varieties in the regions surrounding them. In that case SED maps would be missing islands of highly divergent usage in places where spatial models \cite{bur17,bur18,bur19} suggest that population density gradients should preserve those distinctions or encourage the spread of urban variants. However, the rapid urbanisation of the industrial revolution had started only a couple of generations before the SED speakers were born; less than the three generations required for new dialect formation, according to standard models \cite{trudgill2000}, and certainly not long enough for urban varieties to have substantially diverged from their regional inputs. This is borne out by evidence. For example, Coates \cite{coates2018}, surveying research on traditional Bristol English, notes: ``what can be shown to distinguish Bristolian at all linguistic levels from other dialects of the region is relatively little''. Clearly city varieties at the time of the SED will have had distinguishing features, just as varieties at any point in space differ in some ways from those nearby; but we can assume that they tended to agree in most respects, and so their omission is not likely to be any more problematic than a situation where any other location happened to be lacking in samples.

\subsection{Current hypotheses about linguistic changes}

\label{sec:linguistic background}

The overwhelming story of change in English English dialects over the 20th and early 21st centuries is one of loss of diversity. In the words of Britain \cite{britain2009}: 

\vspace{0.25cm}
\textit{There has been such considerable and ongoing dialect attrition that the language use reported across the country by Ellis's survey of 1889 seems, in many cases and in many places, quite alien to that spoken just over one hundred years later.}
\vspace{0.25cm}

\noindent
In place of smaller traditional dialects we find distinctions at larger, regional levels \cite{britain2002,torgersen2004,tagliamonte2017}, and much research has focused on these \textit{regiolects} and the processes of regional dialect levelling \cite{kerswill2002a,kerswill2002b,kerswill2003,britain2009b,britain2002} which generated them. Reductions in geographical linguistic differences are not only due to convergence to one of several local variants but also to geographically widespread adoption of common innovations \cite{torgersen2004}. These changes are part of a near-universal pattern across the traditional varieties of Europe \cite{auer2005,royneland2010,taeldeman2010}. 

It is believed that interactions between individuals can lead to a decline in linguistic variation via \textit{accommodation}, where conversation partners adjust their speech to better match each other, and by child learners acquiring accommodated forms \cite{trudgill1986,gil07,bab10,doi76,par12}. However, in certain social contexts, children may also learn variants directly from mobile outsiders \cite{trudgill2011a}. The fact that accommodation is mediated through interactions between speakers has lead some linguists to conclude that the primary drivers in the decline of linguistic diversity are travel, commuting and migration \cite{trudgill1986,milroy2002,britain2002,britain2009b,britain2013a,torgersen2004,hinskens2005}. Other potential drivers discussed include changes in social network structure \cite{trudgill1986,mufwene2001,milroy2002,britain2002,torgersen2004,britain2009b,britain2013a}, the age structure of the community \cite{chambers1992a}, the influence of mass media \cite{taeldeman2010,stuart-smith2013}, normative attitudes and education \cite{watt2000,pedersen2005,taeldeman2010} and relatedly the salience and stereotype status of particular variants \cite{kerswill2002,britain2009b}, identity factors \cite{foulkes1999}, the informalisation of public life \cite{taeldeman2010} and socio-economic forces \cite{britain2009b}. Purely linguistic \textit{internal} factors such as structural regularity, functional economy, or naturalness may determine which variants win out in the levelling process \cite{trudgill1986,hinskens2005,britain2002,britain2009b} or they may not be relevant at all \cite{watt2000,torgersen2004}. Arguments have also been made for the central importance of idealogical factors, in particular strong normative attitudes towards the standard, alongside mobility and contact \cite{foulkes1999,pedersen2005,hilton2010}; parallel to these arguments, it has been suggested that strong alignment of speaker identities with the local community may be enough to check the levelling process and so preserve a distinct local variety \cite{taeldeman2010,tagliamonte2017}. 

There are clearly numerous potential drivers of linguistic change, and it is beyond the scope of this work to quantify their relative importance. It may be that driving processes which operate in different ways at small scales (at the level of motivations, interactions and contacts between individuals) yield similar or identical terms in evolution equations for coarse grained population averages, making the task of inferring the importance of individual effects impossible using such models. However, macroscopic models have been used to infer different \textit{classes} of driving mechanism from non-spatial linguistic time series, specifically exogenous (population level) vs endogenous (individual level) drivers \cite{gha14}. We take a similar approach here, but our two classes are spatial processes (migration, movement), and processes which introduce asymmetry with respect to variants (ideology, internal linguistic effects, social prestige, normative bias toward a standard etc.). We do this by defining coarse grained evolution equations which account for both movement and biased copying.  We then fit our model to the initial and final conditions supplied by the SED and EDA data, to determine the relative importance of each process. Since movement is often viewed as the primary driver of change, we view variant symmetric dynamics as our null model, and determine the extent to which it can explain the observed changes, before breaking variant symmetry by allowing for biasing factors.

\section{The model}

We consider a spatial domain divided into $L$ cells, each containing (approximately) $N$ speakers. The centroids of these cells are written $\bv{r}_1,\bv{r}_2, \ldots, \bv{r}_L$. Consider a linguistic variable with $q \in \{1,2, \ldots\}$ variants, and let the relative frequencies with which these variants are used within cell $\bv{r}$ define the frequency vector $\bv{f}(\bv{r}) = (f_1(\bv{r}), \ldots , f_q(\bv{r}))^T \in \Delta^q$, where $\Delta^q$ is the $q$-dimensional simplex and $^T$ denotes transpose. We have suppressed time dependence for brevity. The cells we use in our analysis are Middle Layer Super Output Areas (MSOAs), a set of $L = 7,201$ geographical polygons, each with a similar number of residents, used for reporting census data in England and Wales \cite{ons11}. The mean MSOA population in 2011 was $7,787$, and the modal area of English MSOAs is approximately $0.29 \text{km}^2$, corresponding to a few tens of streets within a densely populated city. Interactions with areas outside the English borders, for which we lack survey data, are neglected. We justify this simplification on the basis that population densities are low at the Scottish and Welsh borders, the English make up the great majority ($\approx 87\%$) of the British mainland population, and Wales and Scotland possess distinctive cultural and linguistic identities. We assume that much longer range interactions, for example from American TV, cinema, and music, will be subsumed into biases in the \textit{learning function} (see section \ref{sec:learn}) which determines the adoption probabilities of different variants by young speakers.  

\subsection{The language community}

\begin{figure}
	\centering
	\includegraphics[width=0.9\linewidth]{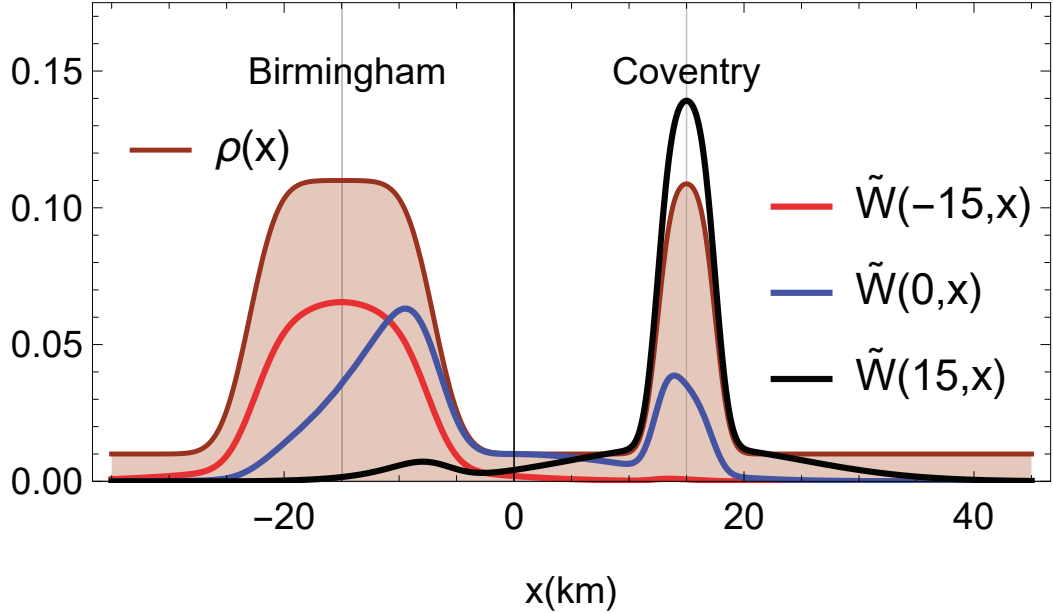}
	\caption{ Effects of towns and cities on the spatial distribution of the 1D continuum influence kernel $\tilde{W}(x_0,x)$, given by (\ref{eqn:W1D}). Brown curve is proportional to population density. Coloured curves show the influences on speakers in the city centres and at the midpoint, when the interaction range is $\sigma=10$km.   }
	\label{fig:birm}
\end{figure}

We refer to the language that a speaker is exposed to as their \textit{linguistic environment}, and assume this environment consists predominantly of voices from their own and nearby cells. We allow for biasing factors by weighting the influence of variants using a set of biases which affect the learning process (see section \ref{sec:learn}). We capture the local environment of a speaker in cell $\bv{r}$ using the \textit{community influence  matrix}, $W$, where $W(\bv{r},\bv{r}')$ gives the influence that speakers from $\bv{r}'$ have on speakers from $\bv{r}$. The matrix is \textit{stochastic}, meaning that it is square with unit row sums. We define the \textit{community frequency vector} of a variable as follows
\begin{equation}
\widehat{\bv{f}}(\bv{r}) \triangleq \sum_{\bv{r}'} W(\bv{r},\bv{r}') \bv{f}(\bv{r}')
\end{equation}
where $\triangleq$ denotes a definition. We assume that local influence strengths are mediated by two factors: physical separation, and population distribution. A definition which incorporates both is 
\begin{equation}
W(\bv{r},\bv{r}') = \frac{\exp \left( -\frac{|\bv{r}-\bv{r}'|^2}{2 \sigma^2} \right)}{\sum_{\bv{r}''} \exp \left( -\frac{|\bv{r}-\bv{r}''|^2}{2 \sigma^2} \right)}, 
\label{eqn:W}
\end{equation}
where we call $\sigma$ the \textit{interaction range}. To illustrate the competing effects of interaction range and population distribution we consider the one dimensional small-cell limit of the influence matrix, which takes the form of a kernel giving the influence of location $x$ on location $x_0$
\begin{equation}
\tilde{W}(x_0,x) = \frac{\rho(x) \exp \left(-\frac{(x-x_0)^2}{2 \sigma^2} \right)}{\int \rho(y) \exp \left(-\frac{(y-x_0)^2}{2 \sigma^2} \right) dy}
\label{eqn:W1D}
\end{equation}
where $\rho(x)$ is population density at $x$, inversely proportional to cell size at this position. Population density is automatically accounted for in definition (\ref{eqn:W}) because higher density areas have more cells. Figure \ref{fig:birm} shows the spatial distribution of influences on three speakers who live in or between two cities with radii and separation similar to Birmingham (population $\approx 10^6$) and Coventry (population $\approx 3 \times 10^5$). 
Consider a speaker in central Coventry ($x_0=15$). Most of the influence on them comes from within the city, because there are relatively few people to interact with outside.  This makes their effective interaction range smaller. A speaker in central Birmingham ($x_0=-15$) will have a wider range of influence because their city is larger. For a speaker in between the two cities ($x_0=0$), influence comes from both, with the larger city dominating (see blue curve in Figure \ref{fig:birm}). In the limit of very large cities (or other regions of approximately uniform population density) the influence distribution becomes normal so that $\approx 63\%$ of influence comes from within $\sigma$km. 

We assume that speakers' linguistic states are fixed in their youth (see section \ref{sec:learn}), so the interaction range should be chosen to capture the community state observed by a younger speaker. An empirical measure of such separations is provided by school travel distances; secondary education has been compulsory since 1918, so we would expect children to regularly interact with others from their school catchment. In the early 21st century secondary pupils travelled, on average, $\approx 5.5$km to school \cite{nts14}. We will call this the \textit{mean catchment radius}, denoted $\mu_R$. The number of state schools has declined somewhat over the 20th century; in 1951 there were 5,900 vs. 4,072 in 2010 \cite{edu12}. This implies that average catchment radii have increased, with 1951 radii being $\approx 83\%$ early 20th century radii. Scaling 2010 travel distances by this factor yields an average travel distance of $\approx 4.5$km. Suppose we model the distribution of displacements between home and school as a bivariate normal distribution
\begin{equation}
f(x,y) = \frac{e^{-\frac{x^2+y^2}{2 \omega^2}}}{2\pi \omega^2},
\end{equation}
then the mean home-school distance is $\mu_R = \sqrt{\tfrac{\pi}{2}} \omega$ and the mean distance between two homes is $2 \omega = \sqrt{\tfrac{8}{\pi}} \mu_R \approx 1.6 \mu_R $. This suggests a typical distance between interacting speakers is between 5 and 10km. Since $\sigma$ functions as an upper bound in cities,  we take our interaction range as the upper limit of this interval, so $\sigma = 10 \text{km} \approx 6 \text{miles}$.

We assume that speakers from a given cell have similar environments, so that $\widehat{\bv{f}}(\bv{r}) $ approximates the perceptions of all speakers in zone $\bv{r}$. The assumption that all the speakers in a cell access the same primary linguistic data is a simplification of reality. We know that within many geographical areas there are subgroups whose language differences may depend on class, sex or ethnicity \cite{cha98}. We simplify in this way because we are interested in national level spatial variations for which we have detailed geographical data, but no consistent breakdown by social subgroup, so cannot infer any effects of subgroups. We implicitly assume that social subgroups can induce biases on variants.

\subsection{Migration}

Migratory processes may be \textit{internal}, where individuals move between locations in a single language area (typically a country), or \textit{external}, where they arrive from another country or language area. During the 20th Century, foreign-born people have made up a small and slowly changing percentage of the English population; 4.2\% in 1951, for example, rising to 6.7\% in 1991 \cite{ons05}. The cumulative total number of migrants to Britain from the early 1800s to 1945 is estimated as $2.34$ million \cite{pan10}. Since numbers are small, we do not explicitly model external migration, instead viewing it as one of many possible factors which may induce biases. To model internal migration, we need to know how often people move, and how far. The English Longitudinal Survey of Ageing (2007), showed that for individuals aged 50-89, the average number of different residences occupied for at least six months during their lifetime was 5.6 for those born 1918-1927, rising to 6.42 for those born 1948-1957 \cite{ber17}, giving an average of approximately one move every ten years. Inter-county migration distances may be extracted from birthplace and place of residence recorded in censuses, providing data back to the mid nineteenth century \cite{fri66}. Such data is, however, spatially coarse; it provides no information about short range migrations. Modern lifestyle survey data can help fill the gap. Analysis \cite{sti16} of large scale research polls by Acxion Ltd (2005-2007), which recorded current and previous postcode, generated $\approx 1.25 \times 10^5$ migration distances, having mean 25.77km, median 2.89km and standard deviation 63.91km. This implies a migration distribution concentrated at short distances, with sub-exponential decline at large distances. This large distance migration behaviour is often modelled with a power, or "gravity" law \cite{sti16,sti78}. 

Models of migration must be consistent with migration distance statistics, and constrained so that total flows in and out of cells match observed values. This can be achieved \cite{wil71} by writing the expected number of people who leave their address in cell $\bv{r}$ and move to a new address in $\bv{r}'$ as
\begin{equation}
m(\bv{r},\bv{r}') \triangleq A(\bv{r}) B(\bv{r}') O(\bv{r}) D(\bv{r}') h(|\bv{r}-\bv{r}'|)
\end{equation}
where $O(\bv{r}) $ is the total number of movers whose previous address was in $\bv{r}$, $D(\bv{r}')$ is the total number of people whose new address is in $\bv{r}'$, and $h(|\bv{r}-\bv{r}'|)$ is a distance function which, typically, decreases for longer moves.  The conditions
\begin{align}
\sum_{\bv{r}'} m(\bv{r},\bv{r}') &=  O(\bv{r})  \\
\sum_{\bv{r}} m(\bv{r},\bv{r}') &=   D(\bv{r}'), 
\end{align}
imply that 
\begin{align}
\label{eqn:A}
A(\bv{r}) &= \left( \sum_{\bv{r}'} B(\bv{r}') D(\bv{r}') h(|\bv{r}-\bv{r}'|) \right)^{-1} \\
B(\bv{r}') &= \left( \sum_{\bv{r}} A(\bv{r}) O(\bv{r}) h(|\bv{r}-\bv{r}'|) \right)^{-1}.
\label{eqn:B}
\end{align}
A mover might shift to a new address in the same cell, and the quantity $m(\bv{r},\bv{r})$ gives the expected number of people who do this within cell $\bv{r}$. The sets of constants $ A \triangleq \{A(\bv{r}_k)\}_{k=1}^L$ and $B \triangleq \{B(\bv{r}_k)\}_{k=1}^L$  are found by generating a sequence of sets $A_0,B_0,A_1,B_1, \ldots$ where $A_0$ is the set for which $A(\bv{r})=1$ for all $\bv{r}$, $B_k$ is computed from $A_k$ using (\ref{eqn:B}), and $A_{k+1}$ is computed from $B_k$ using (\ref{eqn:A}). The sequence is then iterated to convergence.

\begin{figure}
	\centering
	\includegraphics[width=0.7\linewidth]{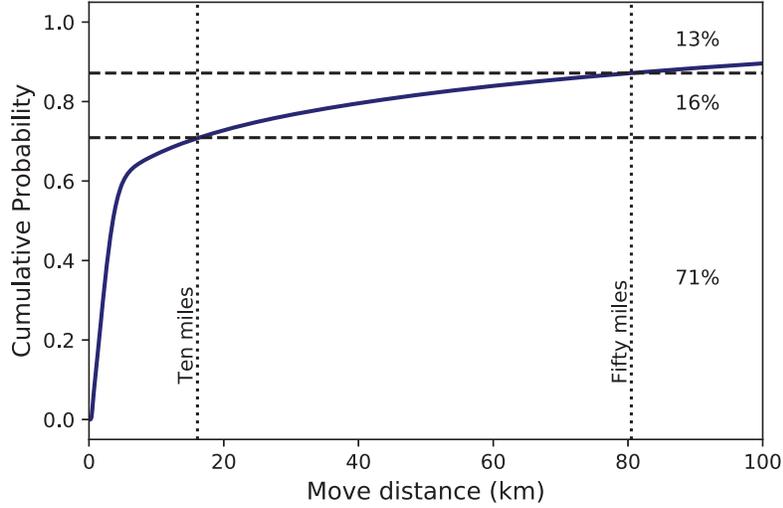}
	\caption{Cumulative distribution of migration distances generated using distance function (\ref{eqn:h}) with $\omega=0.75$, $a=3$km and $\rho=1.5$. The mean, median and standard deviation are 31.3km , 3.6km  and 63.4km  respectively. For comparison, the corresponding statistics computed from Acxion survey data \cite{sti16} are 26.3km, 3.6km, and 63.7km. Horizontal dashed lines show the fraction of moves less then ten miles (71\%) and fifty miles (87\%).  }
	\label{fig:cum}
\end{figure}

We assume that our cell populations are in equilibrium and without loss of generality assume there is one mover per cell, so $O(\bv{r})=D(\bv{r})=1$, and $m(\bv{r},\bv{r}')$ is a stochastic matrix with elements which represent transition probabilities for a single mover. The matrix can be rescaled to capture realistic migration rates. We define the distance function
\begin{equation}
h(r) = \omega e^{-(r/a)^2} + \mathbf{1}_{\{r>0\}}(1-\omega)r^{-\rho}
\label{eqn:h}
\end{equation}
where the first term models short range moves with typical range $a$, and the second term is the long distance fat-tail of the distribution, with gravity exponent $\rho$. The parameter $\omega \in [0,1]$ interpolates between entirely short range, and pure power law distributions. Having calculated $m(\bv{r},\bv{r}')$ for our system, the cumulative distribution of the migration distance $X$ is calculated as
\begin{equation}
\PP(X<r) = \frac{1}{L} \left[\sum_{\underset{\bv{r} \neq \bv{r}'}{\bv{r}, \bv{r}'}} H\left(r-|\bv{r}-\bv{r}'|\right) m(\bv{r},\bv{r}') + \sum_{\bv{r}} H\left(r-R(\bv{r})/\sqrt{2}\right) m(\bv{r},\bv{r}) \right]
\label{eqn:Pcum}
\end{equation}
where $H$ is the Heaviside step function and $R(\bv{r})$ is the radius of a circle with identical area to cell $\bv{r}$. The second term in the square brackets accounts for all intra-cell moves, assumed to be of distance $R(\bv{r})/\sqrt{2}$, following \cite{bat76}. We select distance function parameters to match the short range (median) and long range (standard deviation) statistics of the inter- and intra-cell distance distribution computed in \cite{sti16} for MSOAs, using the Acxion data. This was achieved by parameter grid search to minimize the sum of the relative errors in the median and standard deviation between the model and the Acxion data. The resulting distribution (Figure \ref{fig:cum}) is our estimate for migration distances at the turn of the 21st century. Its shape reflects two kinds of migratory events: frequent short moves, perhaps to change accommodation, and rarer long distance moves, leaving behind friends, work and other social contacts, where the distance travelled becomes relatively less significant.

\begin{figure}
	\centering
	\includegraphics[width=0.7\linewidth]{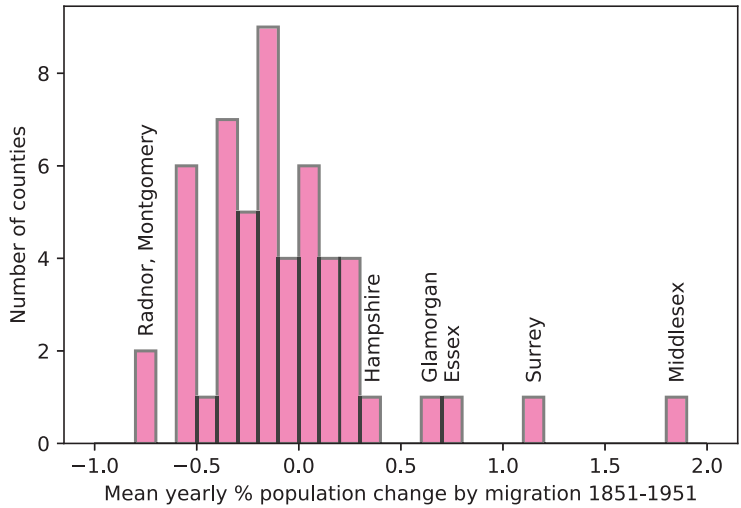}
	\caption{ Yearly percentage changes in population for English and Welsh counties, averaged over 1851-1951, calculated from the census based estimates of Friedlander \cite{fri66}. Counties with the highest and lowest average rates of change are labelled.  }
	\label{fig:net}
\end{figure}

We now consider whether this distribution may be used to approximate migration during in earlier periods. Increased move frequency over the 20th century does not imply that the distribution of distances has also changed. Although we lack detailed information about historical short range moves, we can compare our distribution to statistics of longer migrations extracted \cite{fri66_2} from the National Marriage and Fertility survey (1959-1960), which sampled more than 2000 couples aged 16-60. The survey showed that 16\% of moves were of 10-49 miles and 15\% were 50 miles or more. Our corresponding percentages are 16\% and 13\% respectively (Figure \ref{fig:net}). The similarity of these estimates, and the sample error associated with historical percentages (estimated at $\pm 1\%$), suggests that the statistics of long distance internal migrations have not changed dramatically. Because people are most likely to migrate in their mid twenties \cite{ber17}, many of the moves in the 1959-1960 marriage survey may have been made much earlier in the century. 

We also consider whether our assumption that each cell is in equilibrium is reasonable. Using estimates of net inter-census population changes calculated by Friedlander \cite{fri66} for counties in England and Wales, we have computed the yearly percentage population change by county, averaged over 1851-1951 (Figure \ref{fig:net}). The results show south east England experienced the greatest inflow of migrants, and that the root mean squared net flow rate over all counties is $\approx 0.45\%$. By comparing this figure with the overall yearly migration rate of $\approx 10\%$ we see that, on average, in cells that are gaining, for every 20 people who leave, 21 will arrive. Since net changes in population are therefore typically at least an order of magnitude smaller than than absolute population flows, we assume that non-equilibrium effects can be ignored. 

Long range migration in principle allows variants to jump between locations within the spatial domain. This jumping process is the subject of the linguistic \textit{gravity model} \cite{tru74,cha98}, which attempts to quantify the spread of variants between a finite set of population centres, viewed as points in space. In contrast, the cells in our model cover the entire spatial domain, and population centres appear as densely packed clusters. Other than cell size, there is no intrinsic difference between city and rural cells. The effects of allowing city speakers to behave differently to their rural counterparts, in a spatial model similar to ours, was explored in \cite{bur18}, along with connections to other classical models of spatial linguistic spread including the wave model and hierarchical diffusion \cite{sch72,blo33}.

\subsection{Language learning and evolution}

\label{sec:learn}

We assume an iterative model of language evolution, in which dying adults are replaced by new speakers who select a variant using the probability mass function
\begin{equation}
\bv{p}(\bv{r}) \triangleq \bv{g}\left(\widehat{\bv{f}}(\bv{r})\right) 
\label{eqn:plearn}
\end{equation}
where $\bv{g} : \Delta^q \rightarrow \Delta^q$ is the \textit{learning function}. An alternative interpretation, which yields dynamics which have identical deterministic component, but no stochasticity, is that the components of $\bv{p}(\bv{r})$ give the relative frequencies with which new speakers use different variants. We could, if we wished, interpolate between these two interpretations by assuming a mixture of the two behaviours. In either case equation (\ref{eqn:plearn}) captures the learning process which converts infants into adult speakers. It is an approximation of the cumulative effect of countless small iterations made within a  changing learning environment. Underlying this model is the assumption that linguistic behaviour is mainly acquired in childhood, and changes little in later life. Learning to speak requires the assimilation of a variety of structures, from sets of individual sounds and sound patterns, through words and morphological rules, to syntax. Learning processes differ between structures, and we do not aim to capture these processes in detail. Rather we propose a simple but plausible learning function which maps current speech patterns to learned forms. We derive evolution equations by considering the case which generates maximum stochasticity: where new speakers select a single variant. We will then consider the importance of this stochasticity.

A simple learning model, analogous to \textit{neutral evolution} in genetics \cite{kim71,cox02,kau17,bax06,bly07}, is that variants are selected with a probability equal to their current frequency. This has two problems as a model of language evolution. First, as we will show below, linguistic communities would take an extraordinarily long time to settle on one feature. Second, the geographical boundaries between language features (isoglosses) revealed by linguistic surveys, appear to require some form of social \textit{conformity} \cite{bur17,bur18,bur20} meaning that speakers preferentially select the most common variants with a probability which \textit{exceeds} their relative frequency. Such a selection strategy is optimal if advantage can be gained by matching the speech of others. We can incorporate both conformity and asymmetric bias toward different variants by defining the learning function 
\begin{equation}
\left[ \bv{g}(\bv{f}) \right]_k \triangleq \frac{h_k f_k^\beta}{\sum_{i=1}^q h_i f_i^\beta}
\label{eqn:cm}
\end{equation}
where $[\bv{f}]_k$ denotes the $k$th component of the vector $\bv{f} \in \Delta^q$. The variables $h_k>0$ are biases, and $\beta \geq 1$ is the \textit{conformity number}. If $\beta>1$ then $\bv{g}(\bv{f})$ increases the frequencies of already popular variants, and reduces the frequencies of less popular ones.  The biases $h_1, \ldots, h_q$, which we write as a vector $\bv{h}$, introduce variant asymmetry, with the selection probability of the $k$th variant an increasing function of $h_k$. However, $\bv{g}(\bv{f})$ is invariant under a constant re-scaling of all biases so it is their ratios which determine the extent to which variants are favoured.

To derive equations for the evolution of frequencies, we denote the  counts of speakers using the variants of a given linguistic variable within cell $\bv{r}$ at discrete time step $t$ as
\begin{equation}
\bv{X}(\bv{r},t) = (X_1(\bv{r},t), \ldots, X_q(\bv{r},t))^T
\end{equation}
where
\begin{equation}
\sum_{i=1}^q X_i(\bv{r},t) = N.
\end{equation}
We model \textit{renewal} (birth and death), and \textit{migration}. At each time step, $n_R$ adults die and are replaced with young speakers, and $n_M$ migrate. The renewal and migration rates are 
\begin{align}
\lambda_R & \triangleq \frac{n_R}{N} \\
\lambda_M &  \triangleq \frac{n_M}{N}.
\end{align}
We set $\lambda_M=0.1$, corresponding to one migration per decade, and $\lambda_R=0.02$ corresponding to a typical period of 50 years between reaching linguistic maturity, and leaving the language community. We justify this latter estimate on the basis that linguistic maturity occurs in the late teens \cite{labov2001}, and that UK life expectancy in the mid 20th century was in the late sixties \cite{ons21}. The total number of speakers who die or migrate in each cell is $n \triangleq n_R + n_M$. These speakers are selected uniformly at random from the existing population, implying that both lifetimes and times between migrations are geometrically distributed with parameters $\lambda_R$ and $\lambda_M$, respectively.

If each migrator were to select their new cell using the symmetric transition probability matrix $m(\bv{r},\bv{r}')$, then the expected number of migrators received by each cell would be $n_M$, and the expectation of the mean linguistic state of the migrators incident on cell $\bv{r}$ would be
\begin{equation}
\sum_{\bv{r}'} m(\bv{r}',\bv{r}) \bv{f}(\bv{r}',t) \triangleq \bar{\bv{f}}(\bv{r},t).  
\end{equation}
We model migration by replacing our $n_M$ migrators with $n_M$ speakers with states selected according to the probability vector $\bar{\bv{f}}(\bv{r},t)$. We write the variant counts of these speakers $\bv{X}_M(\bv{r},t) \sim \text{multinomial}(n_M,\bar{\bv{f}}(\bv{r},t))$. The variant counts of the speakers who replace the dead is a random variable, $\bv{X}_R(\bv{r},t) \sim \text{multinomial}(n_R,\bv{p}(\bv{r}))$. The variant counts of the speakers who die or migrate is a random variable, $\bv{Y}(\bv{r},t)$, drawn from the multi-hypergeometric distribution \cite{fel68} with parameters $(n,\bv{X}(\bv{r},t))$. The change in variant counts between $t$ and $t+1$ is then
\begin{equation}
\Delta \bv{X}(\bv{r},t) = \bv{X}_R(\bv{r},t) + \bv{X}_M(\bv{r},t) - \bv{Y}(\bv{r},t) 
\end{equation}
where $\Delta \bv{X}(\bv{r},t) \triangleq \bv{X}(\bv{r},t+1) - \bv{X}(\bv{r},t)$. Dividing by $N$, we have
\begin{align}
\Delta \bv{f}(\bv{r},t) &= \EE[\Delta \bv{f}(\bv{r},t)] + (\Delta \bv{f}(\bv{r},t) - \EE[\Delta \bv{f}(\bv{r},t)]) \\
&= \lambda_R\left(\bv{g}\left(\widehat{\bv{f}}(\bv{r},t)\right) - \bv{f}(\bv{r},t) \right) + \lambda_M \left( \bar{\bv{f}}(\bv{r},t) - \bv{f}(\bv{r},t) \right) + \Delta \bv{Z}(\bv{r})
\label{eqn:df}    
\end{align}
where $\Delta \bv{Z}(\bv{r}) \triangleq \Delta \bv{f}(\bv{r},t) - \EE[\Delta \bv{f}(\bv{r},t)]$ is a zero mean noise term. The variance of the $k$th component of this term, using the properties of the multinomial and multi-hypergeometric distributions, and suppressing $\bv{r}$ dependence for brevity, is
\begin{equation}
\VV[\Delta Z_k] = \frac{1}{N} \left( \lambda_R p_k(1-p_k) + \lambda_M \bar{f}_k(1-\bar{f}_k) + \frac{\lambda(1-\lambda)}{1-N^{-1}} f_k(1-f_k) \right)
\end{equation}
where $\lambda \triangleq \lambda_R + \lambda_M$. The magnitude of the stochastic element of cell dynamics therefore scales as $N^{-1/2}$. For reasons set out below, we assume $\Delta \bv{Z}(\bv{r}) \approx 0$, in the current paper.

\subsection{The importance of noise}

To explore the significance of the noise term in (\ref{eqn:df}), we consider a small, well connected community where linguistic evolution is neutral \cite{kim53,kim71,cox02}, meaning that new speakers adopt variants with probabilities equal to their relative frequencies, so  $\bv{p}(\bv{r})=\bv{f}(\bv{r})$. In the case of a binary variable, the relative frequency of variant 1 admits the diffusion approximation \cite{eth09}
\begin{equation}
df = \sqrt{\frac{2 \lambda_R}{N}} \sqrt{f(1-f)} dW_t
\label{eqn:wf}
\end{equation}
where $W_t$ is a standard Brownian motion \cite{oks10}. Equation (\ref{eqn:wf}) is known as the Wright-Fisher diffusion \cite{eth09} and describes the evolution of the relative frequency of one of two variants (alleles) of a gene in a haploid organism. Notice that the evolution is driven entirely by noise in this case. Let $T$ be the time taken for the cell population to settle on one variant ($f=0$ or $1$). This is the \textit{fixation time}. Starting from the initial condition $f = \tfrac{1}{2}$, we have
\begin{equation}
\EE[T] = \frac{N}{\lambda_R} \ln 2.
\end{equation}
To be concrete, let us consider the case where $N$ is the average population of a single cell ($N=7,787$) in our model, equivalent to a 0.25km$^2$ block of city streets. Using $\lambda_R=0.02$ (as above) we have $\EE[T] = 269,877$ years. From this we see that even in a relatively tiny population, the time for one variant to become dominant by noise alone is comparable to the time for which humans have existed as a species. We therefore neglect the stochastic element of our dynamics on the basis that it is inadequate to explain changes over a plausible time scale. Many of the linguistic variants which interest us have disappeared over the course of a single century and only the deterministic terms in (\ref{eqn:df}) are capable of producing such a change. 

Having decided to neglect the stochastic term in our model, we now consider whether other forms of stochasticity could explain the observed changes. We have assumed our small community is ``well connected'', meaning that all learners are exposed the same inputs. Suppose instead that children learn variants from only a small subset of the community. This effectively subdivides the population into smaller groups (e.g. families) within which variants are reproduced by neutral selection, and between which individuals are exchanged (e.g. by marriage). Such processes are studied in models of neutral genetic evolution in subdivided populations, such Wright's island model \cite{wri43}, and Kimura's stepping stone model \cite{kim53,cox02}. At low exchange rates between groups, the time to fixation increases relative to the fully mixed (single group) case. At sufficiently high mixing rate the population becomes effectively panmictic, and fixation time statistics match the fully mixed case. Subdividing the population therefore does not change our decision to neglect the stochastic term in our model. One way for unbiased random copying to be a more powerful driver of change is via the topology and dynamics of the social contact network \cite{kau17}. For example, if it is strongly clustered then certain individuals or groups become particularly influential. Another possibility is that variants spontaneously gain linguistic \textit{momentum} \cite{sta16,mit11, mich19} starting from small stochastic fluctuations. We do not rule out either possibility, but emphasize that stochasticity as we have modelled it is inconsequential in large populations. This does not mean that we think language change is deterministic. Rather, unpredictable driving forces must result from factors which are not part of our model. Whether their origin is stochastic momentum-like effects, social network dynamics, internal linguistic biases or societal changes outside the linguistic system, they all exert an effective bias (real or apparent) on language change, but in a direction that cannot be known in advance. We can view these factors as random variables whose values are realised as history unfolds. Having observed our linguistic system's historical state, we can then ask: how strong must these biases have been, acting in combination with migration, mixing and social conformity, to produce today's language distributions? Having established which variants require bias in our simple model, we can then turn to understanding their origin.

\subsection{Interpreting model parameters}

Our deterministic evolution equation 
\begin{equation}
\Delta \bv{f}(\bv{r},t) = \lambda_R\left(\bv{g}\left(\widehat{\bv{f}}(\bv{r},t)\right) - \bv{f}(\bv{r},t) \right) + \lambda_M \left( \bar{\bv{f}}(\bv{r},t) - \bv{f}(\bv{r},t) \right) 
\label{eqn:df_det}    
\end{equation}
models four processes, two spatial, two not. The two right hand terms are the \textit{replacement term}, with rate constant $\lambda_R$, and the \textit{migration term} having rate $\lambda_M$.  Mundane mobility is represented by community matrix $W(\bv{r},\bv{r}')$ in our model, which is used to compute $\widehat{\bv{f}}(\bv{r},t)$ and migration by the matrix $m(\bv{r},\bv{r}')$, used to compute $\bar{\bv{f}}(\bv{r},t)$; the effects of these are modulated by conformity $\beta$. Together, these represent a purely spatial, accommodation- and diffusion-driven model which does not differentiate between variants. Factors which might create bias (see section \ref{sec:linguistic background}) are modelled with the bias vector $\bv{h}$. If dialect change has mostly been driven by mobility, as many scholars have suggested, we would expect $\bv{h} \approx \bv{1}$. However, if linguistic factors have been crucial in selecting variants, if the standard ideology has driven levelling, or if local identity factors are important in explaining differences between conservative and innovative regions, then $\bv{h}$ will play a more substantial role. One possibility is that bias drives the selection of the levelling target, but that migration and other spatial factors determine the speed and course of levelling. This would be reflected in the model by high but spatially uniform $\bv{h}$. In such a case, bias might reflect a linguistic constraint derived from a property common to the linguistic system at all localities, or a near-universal social pressure (such as normative pressure from universal schooling). Alternatively, observed regional differences in levelling in the EDA data might not be explained by migration and spatial factors alone and must be modelled with a strongly spatially varying bias. This would imply that local differences in language ideologies play an important role, or that interactions between dialect features create internal pressures varying between dialects.

\section{Interaction between conformity and migration}

In the absence of bias, when $\beta>1$, then the replacement and migration terms in (\ref{eqn:df_det}) represent two competing processes. The replacement term moves the community toward a single local variant, whereas the migration term pulls local forms toward the average linguistic state in a much wider linguistic area. Before comparing our model to data, we illustrate how it operates.

\subsection{Survival of an isolated domain}

\begin{figure}
	\centering
	\includegraphics[width=0.5\linewidth]{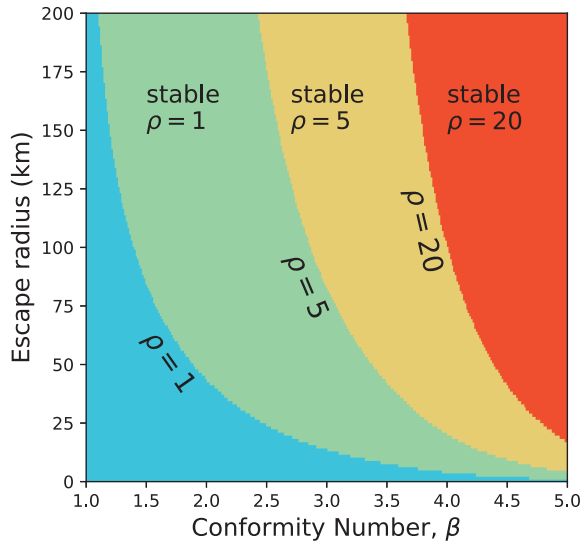}
	\caption{ Phase diagram showing regions of $(\beta,r)$ space for which $\mathfrak{D}$ can retain variant 1 when $\phi_1=0$, for  migration-replacement ratios, $\rho=\lambda_M/\lambda_R$, in the set $\rho \in \{1,5,20\}$. If a region of the phase diagram is stable for some $\rho^\ast$, then it is stable for all $\rho < \rho^\ast$. }
	\label{fig:Phase}
\end{figure}

We consider a binary variable, and let $\bv{f}(t) = (f_1(t), f_2(t))^T$ be its relative frequency vector within a dialect area $\mathfrak{D}$. Let $r$ be the \textit{escape distance}: the expectation of the distance from a randomly chosen point in $\mathfrak{D}$, to the boundary of $\mathfrak{D}$, in a randomly selected direction.  The fraction of migrators who escape $\mathfrak{D}$ is then approximately $1-\PP(X<r) \triangleq \epsilon(r)$ where $\PP$ is the migration measure defined in (\ref{eqn:Pcum}). By the symmetry of the migration matrix, this is equal to the fraction of speakers who enter $\mathfrak{D}$. Writing $\phi_1$ for the relative frequency of variant 1 outside  $\mathfrak{D}$, then the average state of the migrators incident on $\mathfrak{D}$ (including internal movements) will be
\begin{equation}
\bar{f}_1 = \underbrace{(1-\epsilon(r)) f_1}_{\text{moves within $\mathfrak{D}$}} + \underbrace{\epsilon(r) \phi_1}_{\text{migration from outside $\mathfrak{D}$}}.
\end{equation}
The language state of $\mathfrak{D}$ then evolves (assuming $\bv{h}=\bv{1}$) as follows
\begin{equation}
\Delta f_1 = \underbrace{\lambda_R \left[ \frac{f_1^\beta}{f_1^\beta + (1-f_1)^\beta} - f_1 \right]}_{\text{local conformity}} + \underbrace{\epsilon(r) \lambda_M \left[ \phi_1 - f_1 \right]}_{\text{long range levelling}}.
\label{eqn:D}
\end{equation}
The first term generates conformity to the majority variant within $\mathfrak{D}$. The second generates levelling toward the external language state. Suppose that most speakers in $\mathfrak{D}$ use variant 1, which distinguishes them from the nation as a whole, where variant 2 is used (so $\phi_1 \approx 0$). The equilibria ($\Delta f_1 = 0$) of (\ref{eqn:D}) determine the conditions under which $\mathfrak{D}$ will retain its distinctive feature. Informally, we look for the values of $f_1$ for which the rates of conformity and levelling are in balance
\begin{equation}
|\text{local conformity}| = |\text{national levelling}|.
\end{equation}
When $\phi_1=0$, this condition may be written
\begin{equation}
\frac{f_1^\beta}{f_1^\beta + (1-f_1)^\beta} = \left( 1 + \frac{\epsilon(r)\lambda_M}{\lambda_R} \right) f_1.
\label{eqn:lev}
\end{equation}
An upper bound on the migration rate which allows variant 1 to persist within $\mathfrak{D}$ is obtained by considering solutions in the infinite conformity limit $\beta \rar \infty$ where learners always select the majority local variant. In this case the left hand side of (\ref{eqn:lev}) is a unit step at $f_1=\tfrac{1}{2}$, and there is a stable equilibrium $f_1^\ast>\tfrac{1}{2}$ provided
\begin{equation}
\frac{\epsilon(r) \lambda_M}{\lambda_R} <1.
\end{equation}
This is an intuitively reasonable condition; it says that a local distinctive feature can in principle survive provided the rate of inward migration does not exceed the rate at which young indigenous speakers replace dying adults. We estimate that during the 20th century, $\lambda_M \approx 0.1$. There is some flexibility in the way we interpret $\lambda_R$. If we consider speakers to be members of the speech community from birth, then $\lambda_R^{-1}$ is simply life expectancy. However, if speakers are only influential once older, and cease to be so when very old, then $\lambda_R^{-1}$ is the average of this time interval. As noted in section \ref{sec:learn}, we have set $\lambda_R=0.02$. What determines domain stability is not however the absolute values of the migration and replacement rates. It is their ratio
\begin{equation}
\rho \triangleq \frac{\lambda_M}{\lambda_R}.
\end{equation}
For given $\rho$, the stability of $\mathfrak{D}$ can be computed for any value of $\beta$ and $r$ by checking that the dynamics (\ref{eqn:D}) has a stable fixed point $f_1^\ast> \phi_1$. The phase diagram in Figure \ref{fig:Phase} shows that for lower migration-replacement ratios smaller domains can persist for lower values of conformity. In systems where the migration-replacement ratio grows slowly, such as in the English dialect area in previous centuries, we would expect smaller dialect areas to disappear as the phase boundary shifted.

\subsection{Competition between migration and conformity in the English dialect area}

\begin{figure}
	\centering
	\includegraphics[width=0.7\linewidth]{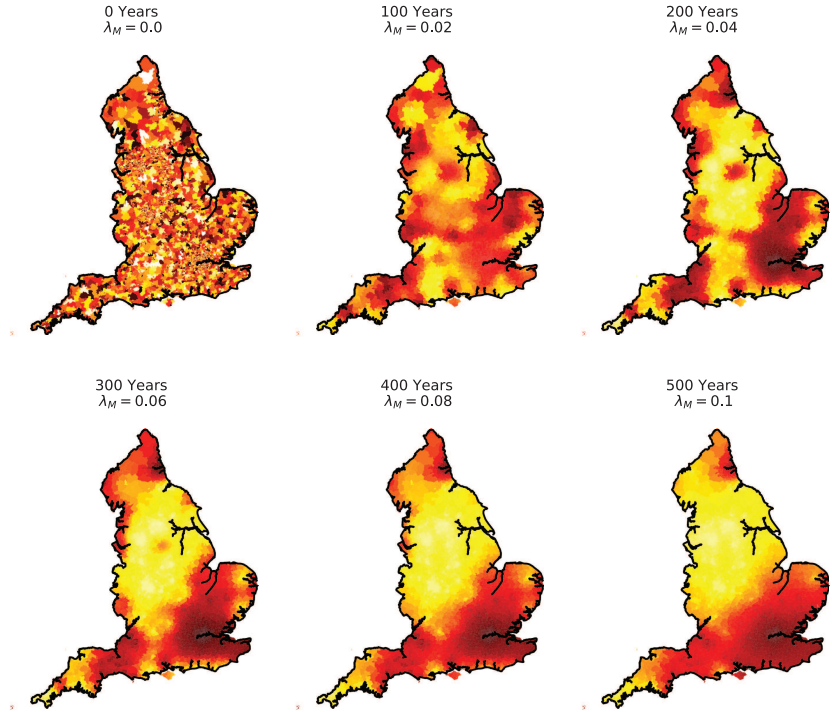}
	\caption{ A 500 year simulation starting from randomized initial conditions with linearly increasing migration rate $\lambda_M(t) = 0.1 \times  t/500$. Other parameters $\lambda_R=0.02, \beta=2.5, \sigma=10$km.}
	\label{fig:domain_evo}
\end{figure}

\begin{figure}
	\centering
	\includegraphics[width=0.7\linewidth]{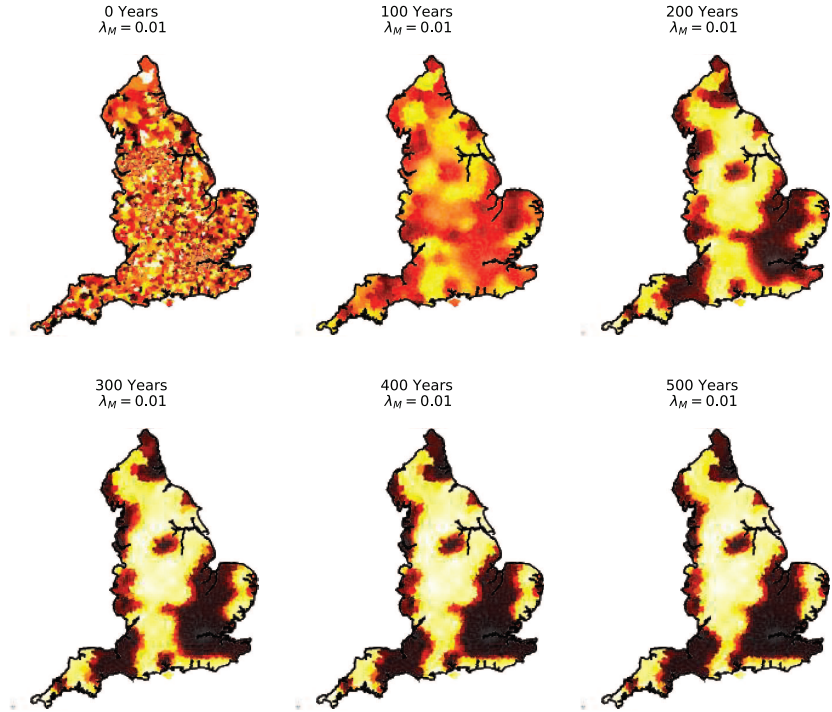}
	\caption{ A 500 year simulation starting from randomized initial conditions with constant low migration rate $\lambda_M = 0.01$. Other parameters $\lambda_R=0.02, \beta=2.5, \sigma=10$km.}
	\label{fig:domain_evo_low}
\end{figure}

We consider a hypothetical period of 500 years, beginning with a domain in a completely randomised state. This state is constructed by setting the initial frequency $f_1(\bv{r},0)$ of each cell to be an independent uniform random variable on $[0,1]$. We allow the migration rate to be a linearly increasing function of time, starting with zero migration and reaching late 20th century levels by end of the period 
\begin{equation}
\lambda_M(t) = 0.1 \times \frac{t}{500}.
\end{equation}
Within a few decades, recognisable dialect domains emerge (Figure \ref{fig:domain_evo}). 
These are analogous to domains in ferromagnetic materials which form because neighbouring atoms tend to align their magnetic directions. Domain boundaries evolve as if they feel a form of surface tension \cite{bra94} which penalises curvature. In physical systems, domains gradually expand as their boundaries straighten. The analogy between these magnetic domain boundaries, and isoglosses, was explored in detail in \cite{bur17,bur18}, including the effects of non-uniform population distribution. It was shown in \cite{bur17} that concentrated human settlements (villages, towns, cities) balance the surface tension effect by inducing isogloss curvature, implying that stable isogloss shapes are partially predictable from population distributions. At low migration levels, competition between isogloss smoothing, induced by surface tension, and curvature, induced by population distribution (as well as other natural barriers) leads to stable geographical distributions. In Figure \ref{fig:domain_evo} the map obtained after 200 years would be stable in the absence of increasing migration. This stability is revealed in Figure \ref{fig:domain_evo_low}, which shows how the system evolves starting from an initial condition which is an exact copy of that in Figure \ref{fig:domain_evo}, but under conditions of consistently low migration ($\rho=\tfrac{1}{2}$). 
This shows that the pattern reached after 200 years remains stable for the next 300 years, and includes a number of isolated dialect areas with small escape radii. Returning to Figure \ref{fig:domain_evo} we see that as migration increases these small dialect areas disappear, being replaced with the variant which is in the majority further afield. Eventually the system reaches a state observed in both the SED and EDA surveys, which consists of a major isogloss bisecting England. The stable north east dialect centred on Tyneside is also seen in  survey data. The major Wash-Severn isogloss is particularly stable because neither variant is in the majority in the country as a whole, and because it bisects the country via one of the shortest routes. A detailed mathematical explanation of these effects, which are related to continuum percolation, is given in \cite{bur17,bur18}. The north east dialect is stable in this simulation because of its status as a dense but geographically isolated population centre.

\section{Inference for English dialects}

\begin{table}
	\caption{\label{tab:vars} \small{Questions, variables and variants for the EDA (and SED). All phonetic variants are written using the International Phonetic Alphabet \cite{internationalphoneticassociation1999}. ME = Middle English} }
	\footnotesize
	\begin{center}
		\begin{tabular}{|p{6cm}|p{4cm}|p{6cm}|}
			\hline
			\small{\textbf{Question in the EDA}} & \small{\textbf{Variable}} & \small{\textbf{Variants}}	\\
			\hline \hline
			1. The word ``tongue'' can end in two different ways. Which do you use? &  Coalescence of [\textipa{Ng}] & 1. [\textipa{Ng}], 2. [\textipa{N}]  \\
			\hline
			2. Which pronunciation of ``new'' is the most similar to your own? & yod-dropping & 1. no [\textipa{j}], 2. [\textipa{j}] \\
			\hline
			3. How do you pronounce	the word ``last''? & TRAP-BATH split & 1. [\textipa{A:}], 2. [\textipa{a:}], [\textipa{\ae:}] 3. [\textipa{a}], [\textipa{\ae}]  \\	
			\hline
			4. In the word	``butter'', I pronounce the letter ``u'' as...& FOOT-STRUT split & 1. [\textipa{U}], 2. [\textipa{2}] \\
			\hline
			5. Do you pronounce the ``r'' in ``arm''? & rhoticity & 1. no /r/, 2. /r/  \\
			\hline
			6. How do you pronounce the word ``three''? & realisation of word-initial /\textipa{T}/ & 1. [\textipa{T}], 2. [\textipa{d}], [\textipa{\:d}], 3. [\textipa{f}], 4. [\textipa{t}]  \\
			\hline
			7. Do you pronounce the ``r''	before the ``-ing'' in ``thawing''? & intrusive /\textipa{r}/ & 1. /\textipa{r}/  2. no /\textipa{r}/  \\
			\hline
			8. What is the season that follows summer? & autumn lexical item & 1. autumn, 2. backend, 3. fall  \\
			\hline
			9. How do you pronounce	the ``l'' in ``shelf''? & l-vocalisation & 1. [\textltilde], 2. [\textipa{u}], [\textipa{U}], 3. [\textipa{l}] \\
			\hline
			10. What do you call a piece of wood stuck under your skin? & splinter lexical item & 1. shiver, 2. sliver, 3. speel, 4. spelk, 5. spell, 6. spile, 7. spill, 8. splint, 9. splinter, 10. spool  \\
			\hline
			11. How do you pronounce the word ``room''? & realisation of ME /\textipa{o:}/ in ``room'' & 1. [\textipa{Y:}], [\textipa{Y}], 2. [\textipa{u:}], 3. [\textipa{U}] \\
			\hline
			12. How do you pronounce the word ``chicken''? &  weak vowel merger & 1. [\textipa{@}], [\textipa{2}], 2. [\textipa{I}] \\
			\hline 
			13. How pronounce the word ``night''? & realisation of  ME /\textipa{ix}/ & 1. [\textipa{AI}], 2.  [\textipa{A:}], 3. [\textipa{O}], [\textipa{6}], 4. [\textipa{Ei}], 5. [\textipa{2i}], [\textipa{@i}], 6. [\textipa{aI}], [\textipa{\ae I}], 7.  [\textipa{a:}], [\textipa{\ae:}], 8. [\textipa{i:}] \\
			\hline
			14. He wasn’t careful with the knife and managed to cut... & initial element in	3sg m. reflexive pronoun & 1. himself, 2.	hisself \\
			\hline
			15. If it belongs to a woman its... & 3sg f. possessive	pronoun & 1. hern, 2. hers \\
			\hline
			16. Do you pronounce the ``h'' in ``hands''? & h-dropping & 1. /\textipa{h}/, 2. no /\textipa{h}/ \\
			\hline
			17. If you’d like someone to pass you something, would you say: & dative alternation & 1. give it me, 2. 	give it to me, 3. give me it \\
			\hline
			18. How do you pronounce the word ``off''? & LOT-CLOTH split  & 1. [\textipa{O:}], 2. [\textipa{A:}], 3. [\textipa{6}] \\
			\hline
			19. How do you pronounce the word ``house''? & realisation of  ME /\textipa{u:}/ & 1. [\textipa{@u}], [\textipa{2u}], [\oe\textipa{7}], 2. [\textipa{E:}], 3. [\textipa{Eu}], 4. [\textipa{aI}], 5. [\textipa{a:}], 6. [\textipa{\ae u}], [\textipa{\ae 7}], 7.  [\textipa{au}], 8. [\textipa{u:}] \\
			\hline
			20. How do you pronounces the word ``bacon''? &  realisation of ME /\textipa{a:}/  & 1. [\textipa{I@}], [\textipa{Ia}], 2. [\textipa{\ae I}], 3. [\textipa{e:}], [\textipa{e@}], 4. [\textipa{ei}], [\textipa{Ei}], [\textipa{E:}], [\textipa{E;@}]  \\
			\hline
			21. How do you pronounce the word ``five''? & realisation of ME /\textipa{i:}/ & 1. [\textipa{AI}], 2. [\textipa{A:}], 3. [\textipa{OI}], 4. [\textipa{EI}], 5. [\textipa{2I}], [\textipa{@I}], 6. [\textipa{aI}], [\textipa{\ae i}], 7. [\textipa{a:}], [\textipa{\ae:}] \\
			\hline
			22. How do you pronounce the last sound of ``bit'' when	saying ``a bit of''? & coda /t/ & 1.[\textipa{P}] , 2. [\textipa{d}], [\textipa{R}] , 3. [\textipa{t}] \\
			\hline 
			23. Fill in the gap: Every day on her walk past the lake, she ... the ducks. & habitual present 3sg. & 1. do feed, 2. feed, 3. feeds \\
			\hline
			24. How do you pronounce the word ``happy''? & happY tensing & 1.  [\textipa{9}], 2. [\textipa{I}], 3. [\textipa{e}], 4. [\textipa{i}] \\
			\hline
			25. An animal that carries its house on	its back is a... & snail	lexical	item & 1. dod-man,  2. hodmedod, hoddy-dod, hoddy-doddy, 3.	snail \\
			\hline	
		\end{tabular}
	\end{center}
\end{table}

\subsection{Methodology}

We now explore how our model predicts the evolution of linguistic variables which have been recorded in both the SED and the EDA. These are listed in Table \ref{tab:vars}. For each variable we evolve our model $T=100$ years forward in time using its SED state as our initial condition. To generate this initial condition from raw survey data, we  set variant frequencies in each model cell (MSOA) to match the variant frequencies from the nearest SED survey location. If multiple variants were recorded in this location, they are assumed to all have equal frequency. This process yields a tessellation of England into single and shared-variant regions. Relative frequencies in each cell are then replaced with average frequencies over its 20 nearest neighbours (mean separation 5.4 km). This amounts to a form of smooth interpolation between relative frequencies observed at SED survey points, which have typical separation $\approx 20$km (calculated as square root of mean land area per location). The resulting distribution has smooth transitions with widths typically 10-30km, consistent with English transition zones discussed in Ref. \cite{cha98}.  While we would not expect our results to be sensitive to the number of nearest neighbours used in the smoothing (interpolation) step, we have avoided smoothing to a level where frequency values at SED locations would be changed substantially in transition regions (e.g. if the smoothing range approached the typical separation of survey locations).

We consider both bias free ($\bv{h}=\bv{1}$) and biased evolution. We allow for the possibility that biases may be different in different parts of the country by introducing a spatially varying field
\begin{equation}
\bv{h}(\bv{r}) = (h_1(\bv{r}), \ldots, h_q(\bv{r}))^T
\end{equation}
where $h_k(\bv{r})$ is the bias on variant $k$ in cell $\bv{r}$.  We do not allow the bias field to depend on time because we have only initial and final conditions, and cannot therefore calibrate time dependence. We wish to fit $\bv{h}(\bv{r})$ so that the final state of our simulation, initialised with an SED variable, closely matches the corresponding EDA data. This is achieved by iterating the following \textit{learning step}, which increases or decreases local biases on variants which are respectively under- or over-represented  in the model, when compared to the EDA
\begin{equation}
\widehat{\bv{h}}_{n+1}(\bv{r}) = \eta ( \bv{f}_{\text{EDA}}(\bv{r}) - \bv{f}_n(\bv{r},T) ) + \theta(\bv{1}-\bv{h}_n(\bv{r})).
\label{eqn:iter}
\end{equation}
Here $\bv{h}_n(\bv{r})$ is the $n$th iteration of our bias estimate and $\bv{f}_n(\bv{r},T)$ is the frequency distribution obtained using this estimate. The \textit{learning rate}, $\eta$, controls how rapidly adjustments are made, and the \textit{reversion rate}, $\theta$, is a regularization parameter that prevents runaway bias increases, and maintains $\bv{h}=\bv{1}$ as the ``no-bias'' reference point so that calibrations for different variables can be compared.  We use the initial condition $\bv{h}_0(\bv{r})=\bv{1}$.

To avoid over-fitting we introduce a smoothing factor, $\sigma_s$, which interpolates between independent variation in bias between cells ($\sigma_s=0$) and constant bias in all cells ($\sigma_x \rar \infty$). The quantity $\widehat{\bv{h}}_{n+1}(\bv{r})$ is the new bias estimate \textit{before} spatial smoothing (indicated by the hat symbol). Having applied the learning step (\ref{eqn:iter}), we apply a \textit{smoothing step} using the interaction matrix (\ref{eqn:W}) with interaction range $\sigma_s$, 
\begin{equation}
\bv{h}_{n+1}(\bv{r}) = \sum_{\bv{r}'} W_{\sigma_s}(\bv{r},\bv{r}') \widehat{\bv{h}}_{n+1}(\bv{r}').
\label{eqn:smooth}
\end{equation}
This two-step iterative estimation method is continued until the bias field stabilizes. The value of $\sigma_s$ is determined by optimising the trade off between model complexity and model error. We measure complexity, $K_N$, as the average $\ell_1$ norm of the difference between the bias in each cell, and the average, $\overline{\bv{h}}_N(\bv{r})$, of its $N$ nearest neighbours
\begin{equation}
K_N =  \sum_{\bv{r}} \lVert \bv{h}(\bv{r}) - \overline{\bv{h}}_N(\bv{r}) \rVert_1.
\end{equation}
This measure approximates the magnitude of the second spatial derivative of the bias field, averaged over the whole system. The number of neighbour cells ($N$) is chosen to be large enough to smooth out inhomogeneity in the pattern of cell centroids, but small enough so that shorter range fluctuations in the bias field are not ignored. We take $N=20$, corresponding to an average neighbour distance of 5.4 km, on the assumption that repeated application of the smoothing step will ensure that fluctuations on shorter scales are small. Complexity will be high for fields with large variations in bias over short distances throughout the system, and low for fields which are approximately constant over large geographical areas. Higher learning rates will generate bigger local changes at each step, but these are counteracted by increased smoothing range. Reducing the learning rate to $\eta'=\eta/k$, $k \in \mathbb{N}$, is approximately equivalent to applying $k$ smoothing steps between each learning step, or increasing the smoothing parameter to $\sigma_s' = \sqrt{k} \sigma_s$, so we can explore the behaviour of our calibration method by fixing $\eta$ and varying $\sigma_s$. 
\begin{figure}
	\centering
	\includegraphics[width=0.8\linewidth]{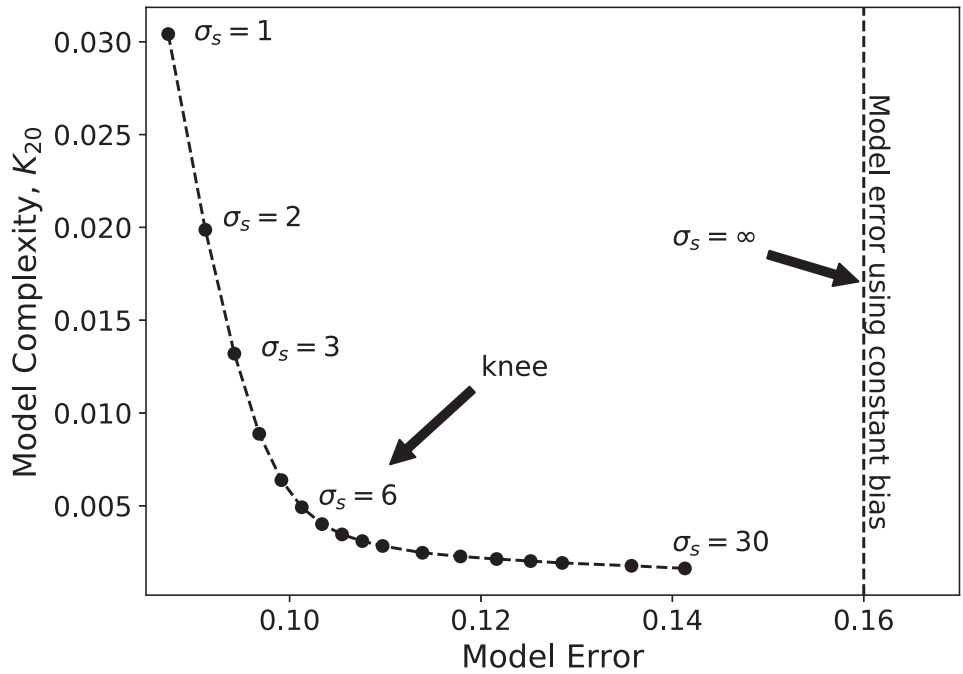}
	\caption{ Model complexity versus model error for variable 7, using learning parameters $\eta=0.15, \theta=0.03$ and smoothing parameters in the range $\sigma_s \in \{1,2,\ldots,10,12,\ldots,20,25,30\}$. Model parameters $\lambda_M = 0.1, \lambda_R=0.02, \beta=2.0, \sigma=10$km.   }
	\label{fig:occam}
\end{figure}
Figure \ref{fig:occam} shows the relationship between model complexity and model error using variable 7 as an example. For high smoothing parameter ($\sigma_s=30$km) the error approaches that of the maximally parsimonious model, where bias is constant in space. Reducing $\sigma_s$ improves the model error, until we reach an optimal region ($\sigma_s \approx 6$km), the ``knee'' point, with low model errors and minimal complexity. Further increasing complexity has minimal effect on error. We take this knee as our heuristic for determining the optimal level of complexity because this is the simplest explanation that fits the data (Occam's Razor \cite{mac03}). Exploration of the error vs. complexity trade-off for other questions reveals similarly located knee points. For consistency we use $\sigma_s=6$km for all questions (results with $\sigma_s=10$km are shown in supplementary material \cite{bur21}). For comparison, we also calibrate with very high smoothing $\sigma_s=150$km in the case $\beta=2$, which produces an approximately constant bias field.  We select learning parameters $\eta=0.15, \theta=0.03$ to allow convergence within a feasible time frame, while maintaining stability in the iteration. 

We measure the difference between modelled and observed variant distributions at time $T$ using the \textit{total variation distance} \cite{csi04} between the variant frequency distributions in each cell, averaged over all cells
\begin{equation}
\text{error} \triangleq \frac{1}{2L} \sum_{\bv{r}} \lVert \bv{f}_{\text{model}}(\bv{r})-\bv{f}_{\text{EDA}}(\bv{r})\rVert_1 \in [0,1]
\label{eqn:err}
\end{equation}
where $\lVert \cdot \rVert_1$ is the $\ell_1$ norm. The factor $(2L)^{-1}$ ensures that the maximum error is 1. In this extreme case no speaker in any cell uses a variant observed in that cell in the survey.

The model is implemented in vectorized Python using sparse arrays which accelerate the calculation of the community frequency vector. The number of bias field iterations is set between 25 and 35, sufficient for the error (\ref{eqn:err}) to stabilize. To calibrate a single question using a single set of model parameters takes approximately 5-10 minutes on a modest laptop (1.7 GHz, 4 cores). For much larger systems than ours, there is potential for the migration matrix to create a bottleneck, requiring further optimization.

\subsection{Bias free results}

\begin{figure}
	\centering
	\includegraphics[width=\linewidth]{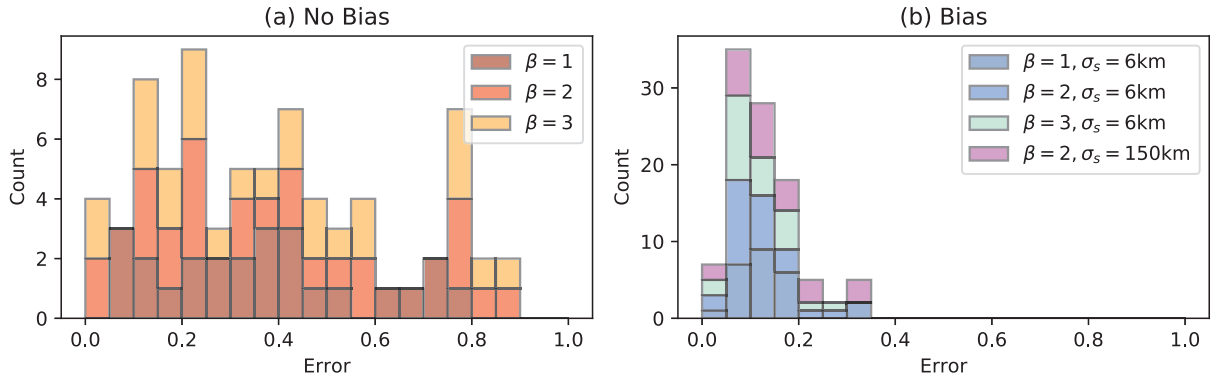}
	\caption{Stacked histograms of model errors over all variables for conformity $\beta \in \{1,2,3\}$, $\lambda_R=0.02, \lambda_M=0.1$. (a) Errors in the bias free case $\bv{h}=\bv{1}$. (b) Errors after calibrating bias using (\ref{eqn:iter}) and (\ref{eqn:smooth}). Very high smoothing parameter $\sigma_s=150$km leads to approximately constant spatial bias.  }
	\label{fig:errs}
\end{figure}

We begin by considering the bias free case ($\bv{h}=\bv{1}$) where the only factors driving evolution are migration, daily movement and conformity. Figure \ref{fig:errs} (a) shows the range of model errors in this case. In approximately $25\%$ of cases the model generates a final distribution with error ($< 0.2$) comparable with errors obtained by fitting a bias field  (Figure \ref{fig:errs} (b)). These are variants for which spatial process and conformity play a substantial role in explaining their evolution. Many exhibit one of two phenomena seen in two dimensional coarsening systems studied by physicists \cite{bra94}: stripe states, and shrinking droplets.  

\subsubsection{Stripe states}

\begin{figure}
	\centering
	\includegraphics[width=\linewidth]{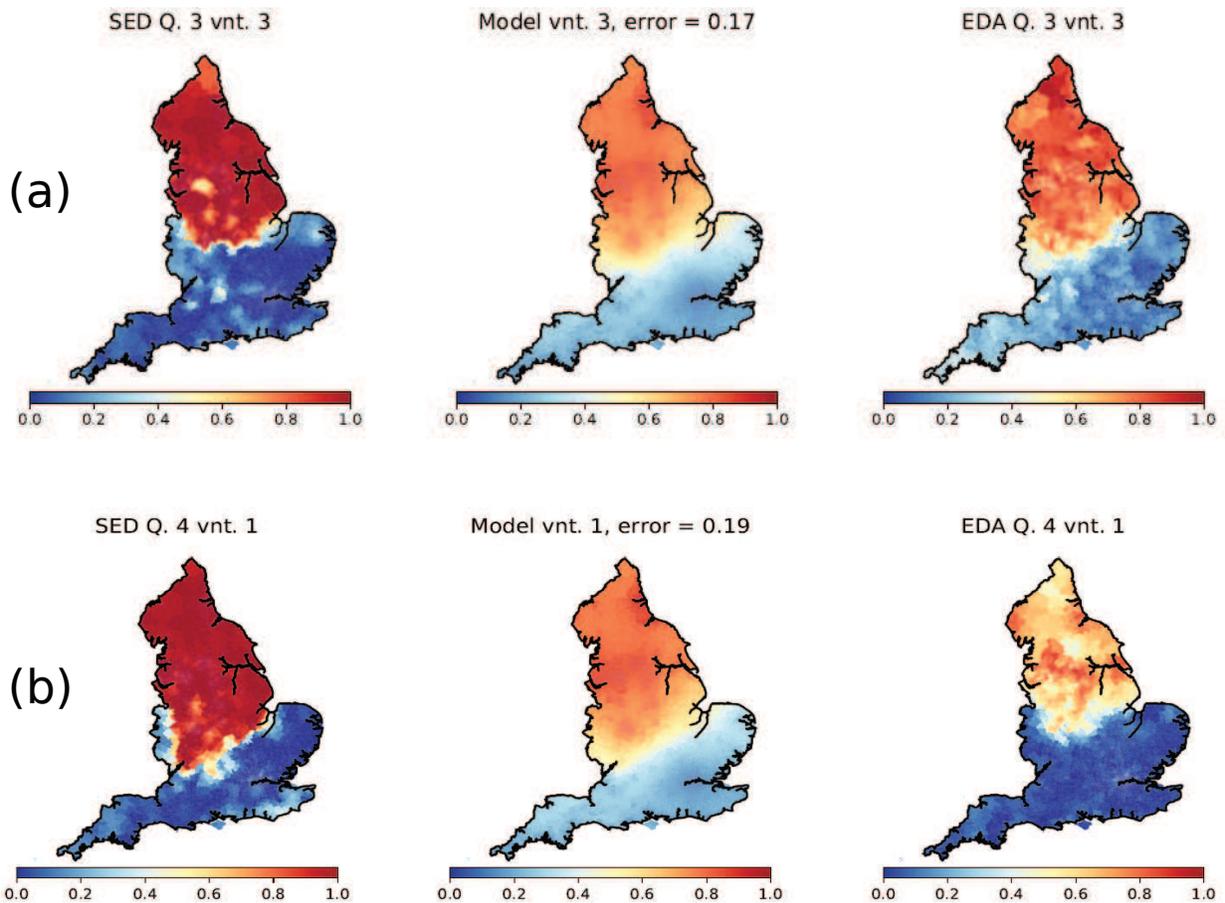}
	\caption{ SED (left),  no-bias model results at $T=100$ (middle) and EDA (right) distributions for (a) Variant \textipa{[a]} of variable 3, the \textsc{trap-bath} split,  (b) Variant \textipa{[U]} of variable 4, the \textsc{foot-strut} split. Model parameters $\lambda_M = 0.1, \lambda_R=0.02, \beta=2.0$.  }
	\label{fig:q3_4_no_bias}
\end{figure}

In the bias free case, provided $\beta>1$, our model behaves like a subcritical coarse grained two-dimensional ferromagnet \cite{bra94, bar09}. The condition $\beta>1$ allows interfaces to form between single variant domains, analogous to domains of uniform spin alignment in magnetic systems. These interfaces feel a form of surface tension, and in finite systems, stable configurations often emerge where interfaces connect opposite system boundaries \cite{bar09}, creating \textit{stripes}. Variant \textipa{[a]} of variable 3 in Table \ref{tab:vars} illustrates this phenomenon (Figure \ref{fig:q3_4_no_bias} (a)).   
Here, we see that the geographical boundary between those who pronounce the vowel in the \textsc{bath} lexical set \cite{wells1982} (\textit{last}, \textit{staff}, \textit{brass}, \ldots) in the same way as vowel as the \textsc{trap} set (\textit{tap}, \textit{back}, \textit{badge}, \ldots), from those who do not, has been stable over the 20th Century. Note that \textit{lexical sets} (intoduced by John C. Wells \cite{wells1982}) are groups of words defined by some common phonological feature. The variant  \textipa{[a]} (the vowel in northern ``bath'') dominates in the northern region, where the vowels in \textsc{trap} and \textsc{bath} do not differ. This same stability is predicted by the model, and derives from the surface tension effect which causes interfaces to find the shortest routes across the system, which often begin in boundary indentations \cite{bur17} (the Wash in Figures \ref{fig:q3_4_no_bias} (a) and (b)). One important difference between our model and simple magnetic systems is migration, which induces effective long-range interactions equivalent to local biases in favour of the national majority variant. Because the \textsc{trap-bath} split divides the country into approximately equal parts, these biases are close to zero, stabilizing the boundary. 

Another example of this effect, shown in Figure \ref{fig:q3_4_no_bias} (b), is the distribution of the variant \textipa{[U]} of variable 4. This is the northern English form of the stressed vowel in ``butter'' which belongs to the Wells' lexical set \textsc{strut} (\textit{cup}, \textit{suck}, \ldots). Speakers from the north typically use the same vowel sound for the lexical set \textsc{foot} (\textit{put}, \textit{bush}). Although the no-bias model captures the broad overall pattern of \textsc{foot-strut}, we can see from the EDA distribution that the northern form is in fact receding, which indicates a bias in favour of the southern pronunciation, \textipa{[2]}. Further evidence of such a bias, is that the model predicts levelling of both variants, when in fact the southern form remains dominant in the EDA.

The effect of increasing the conformity number is to make single variant regions increasingly pure, boundaries sharper, and more strongly affected by variations in population distribution.  In this paper, for every variable, we have run the model with $\beta \in \{1,2,3\}$.  Mean errors in the biased and no-bias model are summarized in Table \ref{tab:err}. For the no-bias model we compute average errors excluding cases where these exceed a given threshold, $C$, allowing us to consider only cases where the model provides a plausible description of changes. In the biased model we find that conformity generates a better fit than no conformity (a paired t-test with null hypothesis of equal mean error between $\beta=1$ and $\beta=2$ yields p-value 0.0016). There are small differences ($ \lessapprox 0.01$)  between the cases $\beta=2$ and $\beta=3$. In the no-bias model, when errors are within one standard deviation of the biased model error, the conformity model performs marginally better than the conformity free case. If we raise the error threshold to two standard deviations the results are equivocal. Table \ref{tab:err} therefore provides only limited evidence that $\beta>1$ in the no-bias case. We suggest that $\beta>1$ is more plausible because in the long term, linguistic boundaries could not spontaneously emerge or be maintained if $\beta=1$. Moreover, without conformity the unbiased dynamics simply mixes variants, dissolving boundaries, but leaving total variant frequencies over the system as a whole unchanged, meaning that levelling is impossible. Previous modelling work \cite{bur20} also indicates that the SED distributions are more likely to have been generated by a model with conformity. Finally, we note that the substantial linguistic changes observed during our period of interest may be driven by a wide range of factors, of which conformity is just one, makings its effects harder to discern. A complete set of maps for all variants of all variables for the SED, EDA, no-bias, and bias models (bias fields included) for $\beta=2$, is given in the supplemental material for this paper. For the remainder of the paper we focus on the case $\beta=2$, and consider spatial varying bias ($\sigma_s=6$km), approximately constant bias ($\sigma_s=150$km) and no bias.

\begin{table}
	\caption{\label{tab:err} Mean errors for $\beta \in \{1,2,3\}$. Here $\varepsilon_{\text{bias}}$ and $\varepsilon$  denote errors in the bias and no-bias models respectively. The shorthand  $\la \varepsilon|\varepsilon < C \ra$ denotes mean error conditional on $\varepsilon<C$. Counts give the numbers of questions satisfying the condition. Note that $C=0.17$ is one standard deviation above the mean error in the $\beta=2$ bias model, and $C=0.22$ is two standard deviations above. For comparison, the mean error with $\beta=2$ and spatially constant bias is $0.15 \pm 0.09$.}
	\begin{center}
		\begin{tabular}{|c|c|c|c|c|c|}
			\hline
		$\beta$ & $\la \varepsilon_{\text{bias}} \ra \pm \text{std. dev.} $ & $\la \varepsilon|\varepsilon<0.17 \ra$  & Count($\varepsilon<0.17$) & $\la \varepsilon|\varepsilon<0.22 \ra $ & Count($\varepsilon<0.22$)	\\		
		\hline
		\hline
		1 & $0.133 \pm 0.070$ & 0.105 & 6 & 0.121 & 7 \\
		2 & $0.109 \pm 0.058$ & 0.087 & 5 & 0.123 & 8\\
		3 & $0.112 \pm 0.061$ & 0.089 & 5 & 0.126 & 8\\
		\hline
	\end{tabular}
\end{center}
\end{table}

\subsubsection{Shrinking droplets}

\begin{figure}
	\centering
	\includegraphics[width=\linewidth]{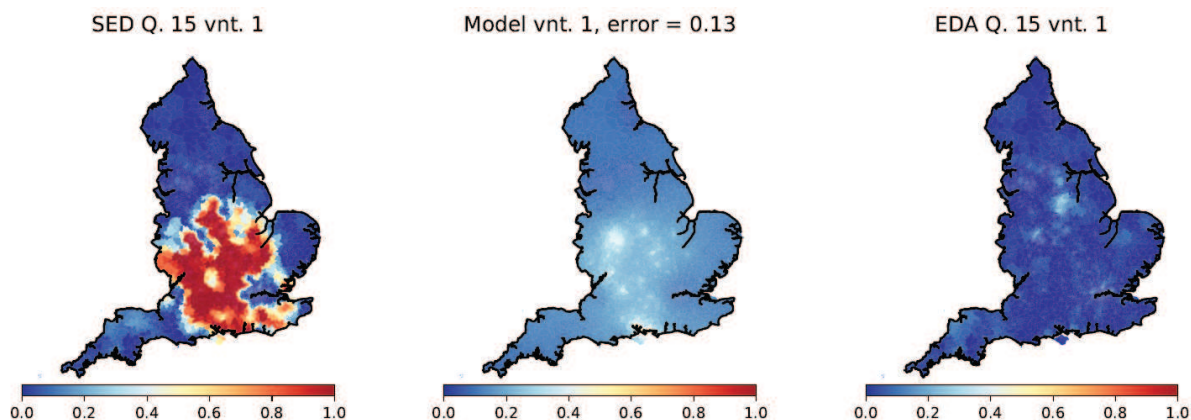}
	\caption{ SED (left),  no-bias model results (middle) and EDA (right) distributions for spatial distribution of variant 1 (\textit{hern}) of SED variable 15. Model parameters $\lambda_M = 0.1, \lambda_R=0.02, \beta=2.0, \sigma=10$km. Model error = 0.13.}
	\label{fig:Q15_beta2}
\end{figure}

Another phenomenon seen in ferromagnetic interfaces, and in our model, is the \textit{shrinking droplet}. In two dimensions a droplet is simply a closed curve, evolving under surface tension, which acts to shorten its length. As a result, droplets become more circular and eventually shrink to nothing. In our model, droplets of variants which are in the minority are further reduced by migration. As an example, consider the form of the 3sg feminine possessive pronoun, \textit{hers} or  \textit{hern} (variable 15, Figure \ref{fig:Q15_beta2}).
\textit{Hern} is a minority variant in the SED (29\% vs. 71\% nationally), surrounded by regions in which \textit{hers} is dominant. Without conformity, the mixing effect of migration would move the system as a whole to a spatially uniform state with all cells having frequency vector $\bv{f}=(0.29,0.71)$. With conformity, surface tension acting at the boundary of the hern region, plus the effects of migration, almost extinguish the \textit{hern} variant within 100 years. Examples of droplet shrinkage from a different class of linguistic feature are found in the reflexes of Middle English \textipa{/u:/} (variable 19), where minority variants such as \textipa{[E:]} and \textipa{[aI]} are predicted to undergo near-complete levelling (disappearance) without bias.

\subsection{Bias driven evolution}

\begin{figure}
	\centering
	\includegraphics[width=\linewidth]{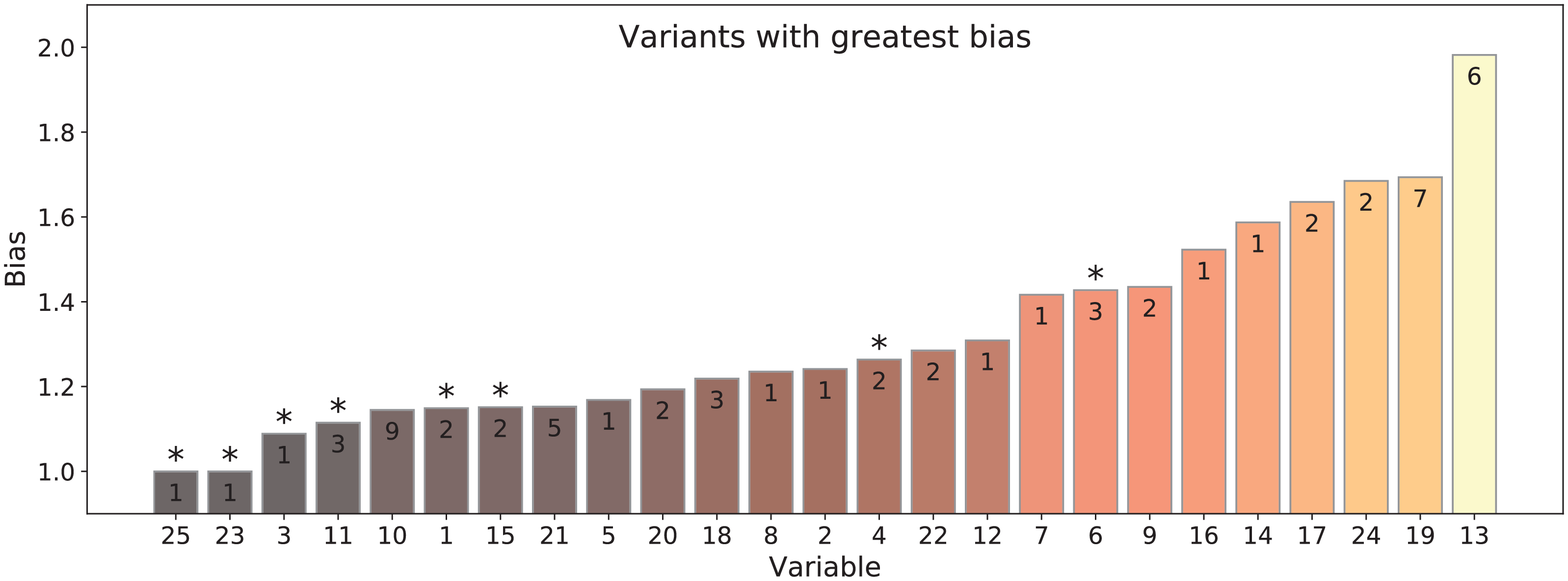}
	\caption{Variants with the maximum average bias per cell (given by bar height) in the bias model with parameters $\lambda_M = 0.1, \lambda_R=0.02, \beta=2.0$. Each bar is labelled with the variant that experiences the maximum bias. Starred bars indicate variables for which the no-bias model with $\beta=2$ generates an error $<0.2$.}
	\label{fig:bias_hist}
\end{figure}

\begin{figure}
	\centering
	\includegraphics[width=\linewidth]{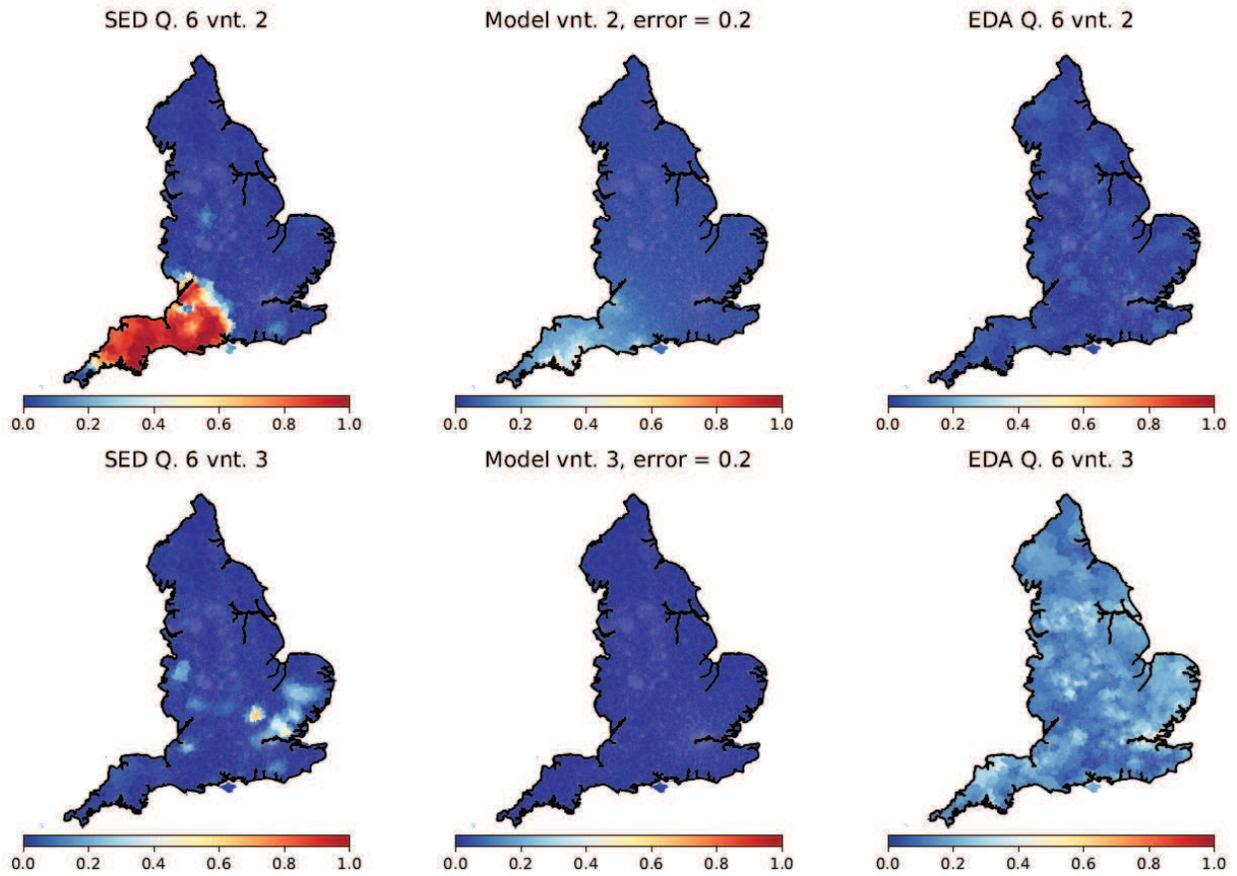}
	\caption{SED (left),  no-bias model results (middle) and EDA (right) distributions for variant 2 ([\textipa{d\textturnr i:}]) and variant 3 ([\textipa{f\textturnr i:}]) of variable 6: pronunciation of ``three''. Model parameters $\lambda_M = 0.1, \lambda_R=0.02, \beta=2.0$.}
	\label{fig:Q6}
\end{figure}

The evolution of many variables in Table \ref{tab:vars} may be understood as the result of combined spatial and bias effects. The primary effect of bias is to lend certain variants a special status which may have either linguistic (\textit{internal}) or social (\textit{external}) origin. In some cases, a bias appears to accelerate primarily spatial processes. Figure \ref{fig:bias_hist} shows the average bias per cell applied to the most biased variant of each variable, using $\beta=2$ (similar results are obtained with $\beta=3$). Variables for which the no-bias model produces an error $<0.2$ are starred and, with two exceptions, these variables have the smallest maximum bias per cell. Bias magnitude therefore provides an indication of the extent to which the pure spatial model fails to explain observed changes between the SED and EDA. For the more extreme of the two exceptions (variable 6, Figure \ref{fig:Q6}), the no-bias model correctly predicts the main 20th century change: disappearance of the stopped pronunciation of ``three'' from the West-Country, but fails to predict that the highly localised Essex pronunciation ``free'', although it remained a minority variant, would spread across the country, requiring a substantial bias in its favour, to prevent complete levelling.      

\subsubsection{Special status variants}

\begin{figure}
	\centering
	\includegraphics[width=\linewidth]{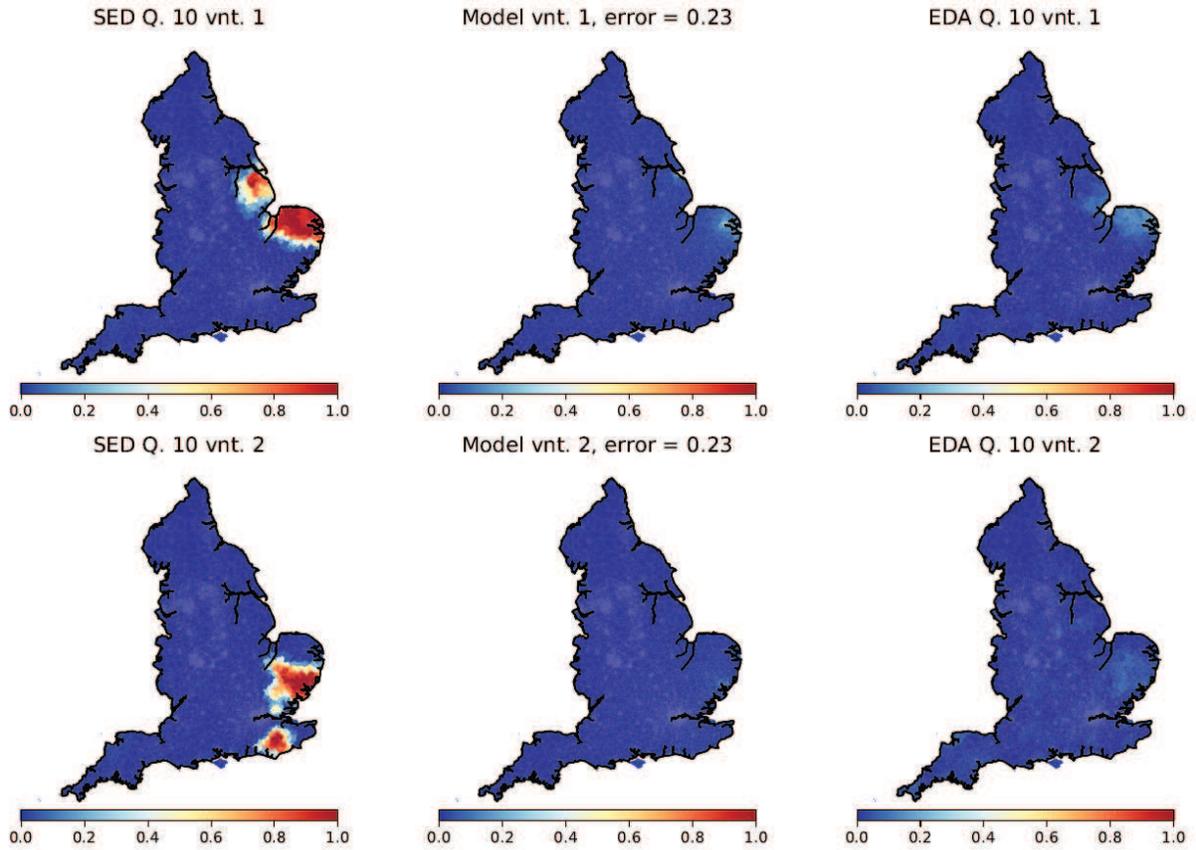}
	\caption{SED (left),  no-bias model results (middle) and EDA (right) distributions for variants \textit{shiver} and \textit{sliver} of variable 10, \textit{splinter}. Model parameters $\lambda_M = 0.1, \lambda_R=0.02, \beta=2.0$.}
	\label{fig:q10_no_bias_results}
\end{figure}

\begin{figure}
	\centering
	\includegraphics[width=\linewidth]{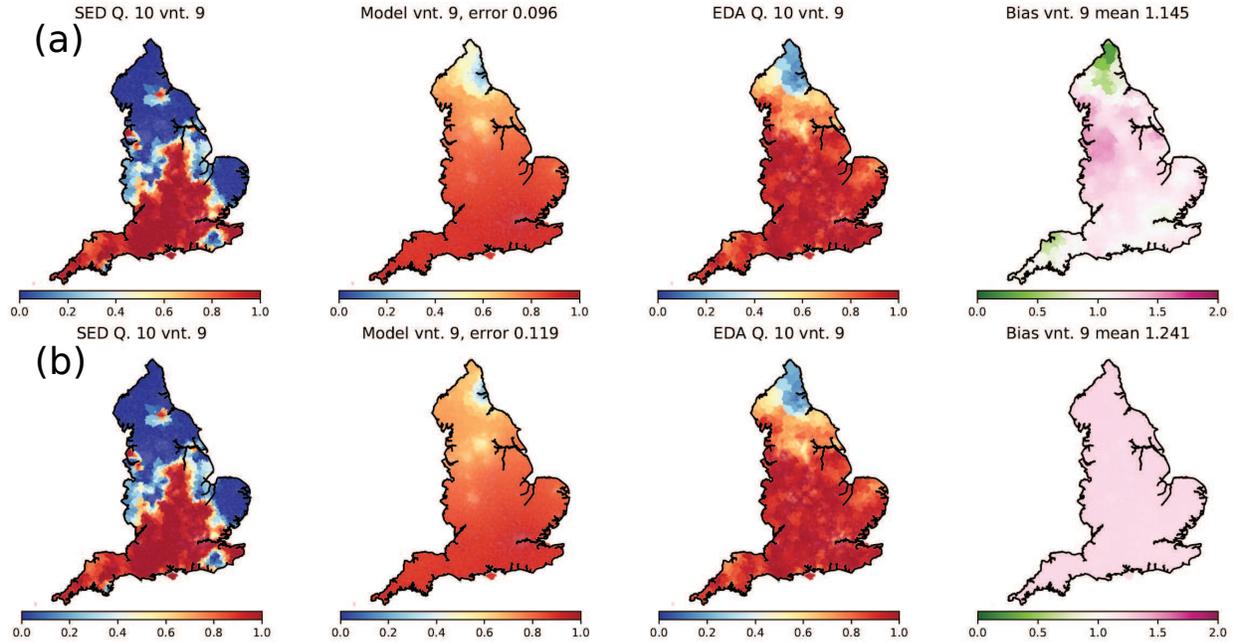}
	\caption{(a) SED (far left),  bias model results (middle left) and EDA (middle right) distributions for variant \textit{splinter} of variable 10. Far right map shows calibrated bias field using $\sigma_s=6$km. Model parameters $\lambda_M = 0.1, \lambda_R=0.02, \beta=2.0$. (b) As for (a), but using $\sigma_s=150$km to yield constant bias field.}
	\label{fig:splinter}
\end{figure}

\begin{figure}
	\centering
	\includegraphics[width=\linewidth]{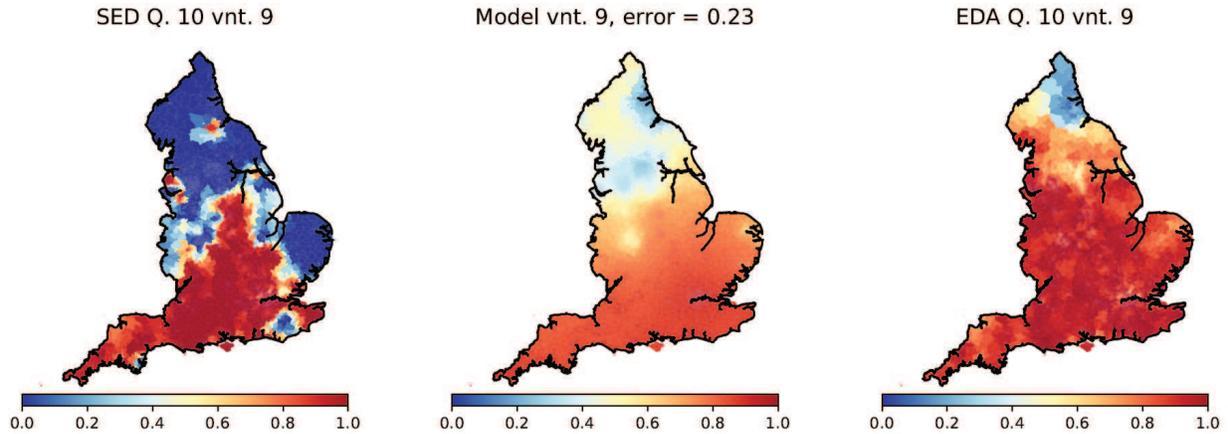}
	\caption{SED (left),  no-bias model results (middle) and EDA (right) distributions for variant \textit{splinter} of variable 10, with model parameters $\lambda_M = 0.1, \lambda_R=0.02, \beta=2.0$. }
	\label{fig:splinter_no_bias}
\end{figure}

Consider variable 10, the word for a small piece of wood stuck under the skin. Minority variants like \textit{shiver} and \textit{sliver} (see Figure \ref{fig:q10_no_bias_results}) are correctly predicted by the no-bias model to undergo near-total levelling.
Such patterns are most typical of variables with numerous variants, many of which were historically restricted to small regions.  In contrast, the present day distribution of \textit{splinter} cannot be explained without bias (Figure \ref{fig:splinter}). In the SED this was the dominant southern variant, and the calibrated bias field indicates that it must have attained a special status as a standard variant in order to spread north. However, this progress was halted in far north east England, where \textit{spelk} persists. Our spatially invariant bias field (Figure \ref{fig:splinter} (b)) indicates that this effect can be partially, but not completely explained by the region's geographical isolation. According to the spatially varying bias field, \textit{spelk} not \textit{splinter} is the variant with special status in the north east (Figure \ref{fig:splinter} (a)), without which it would be dominated by \textit{splinter}, except in the major population centre of the region (Newcastle). A similar effect is observed in variable 12, with the preservation of weak-vowel merger in the north east, when the rest of England has moved to the standard variant (no merger).
The no bias model predicts the initial expansion of \textit{splinter} due to its dominance in the densely populated south. However, its progress north is arrested by repulsion  effects \cite{bur17} created by the major urban conurbations of Manchester, Sheffield and Leeds (Figure \ref{fig:splinter_no_bias}). We find that these cities play an important role in the pronunciation of ``thawing'' (variable 7, Figure \ref{fig:q7_v1_bias_results}). 

An important question is: how should the bias fields be interpreted? In the strictest sense the fields are merely model fitting parameters which measure how much asymmetry needs to be injected into the model to explain the observed changes. Where the variant symmetric model adequately explains the observed changes without the need for additional bias, we expect to see a calibrated bias field close to one. This does not mean that in reality there is no linguistic or social bias, only that conformity and migration alone are sufficient to explain the changes. For example, in Figures \ref{fig:splinter} and \ref{fig:splinter_no_bias} because the no bias model predicts the first part of the northerly expansion of splinter, minimal bias is needed in the south. The major obstacles to further expansion are the cities of Manchester and Sheffield, which consequently have high bias field. This does not imply that surrounding areas necessarily have lower bias, only that it is not needed to explain observations. The bias field is therefore most informative when viewed in conjunction with the predictions of the no-bias model (a complete catalogue is in the supplemental material \cite{bur21}).

\begin{figure}
	\centering
	\includegraphics[width=\linewidth]{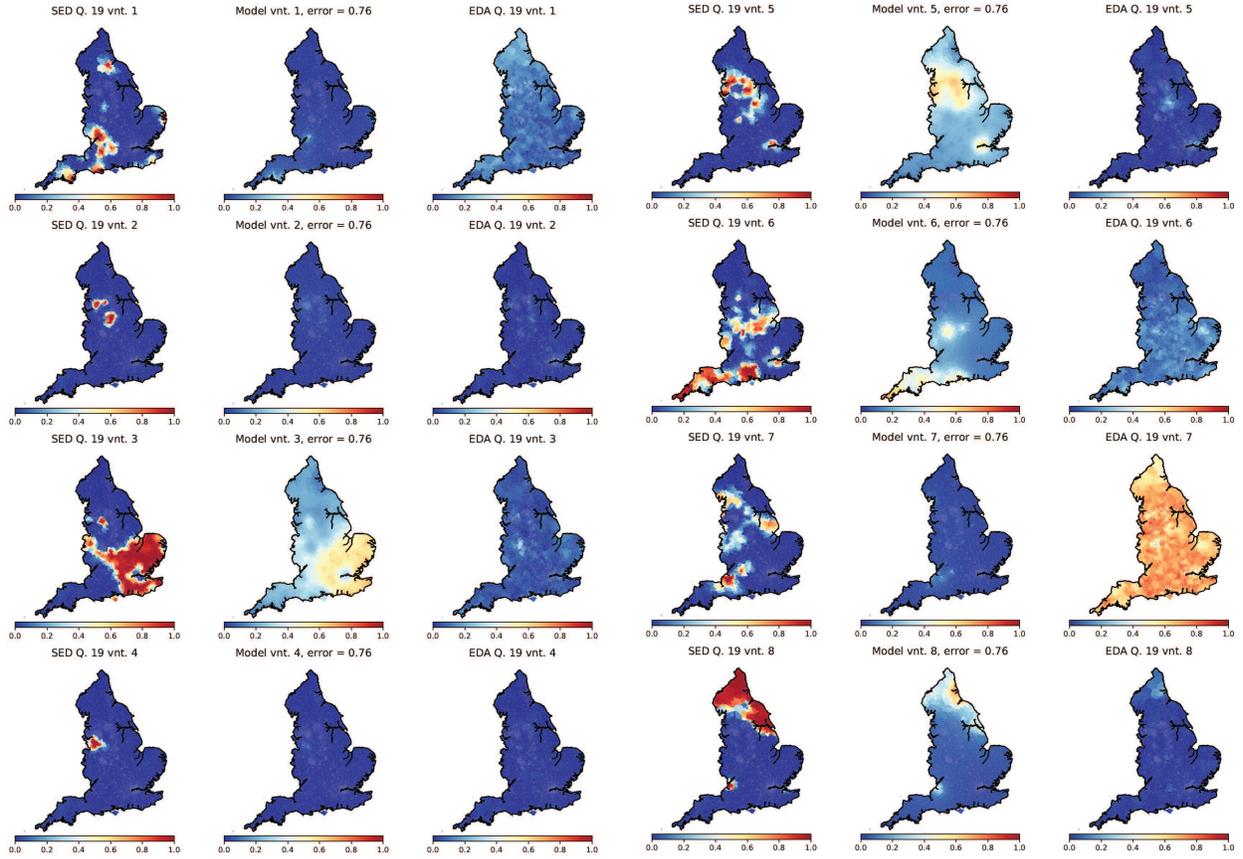}
	\caption{All variants of variable 19: reflexes of ME \textipa{/u:/} in the no-bias model. For each variant we have SED (left),  no-bias model results (middle) and EDA (right) distributions. For reference, variants are 1. [\textipa{@u}], [\textipa{2u}], [\oe\textipa{7}], 2. [\textipa{E:}], 3. [\textipa{Eu}], 4. [\textipa{aI}], 5. [\textipa{a:}], 6. [\textipa{\ae u}], [\textipa{\ae 7}], 7.  [\textipa{au}], 8. [\textipa{u:}]. Model parameters $\lambda_M = 0.1, \lambda_R=0.02, \beta=2.0$.  }
	\label{fig:q19}
\end{figure}

\begin{figure}
	\centering
	\includegraphics[width=\linewidth]{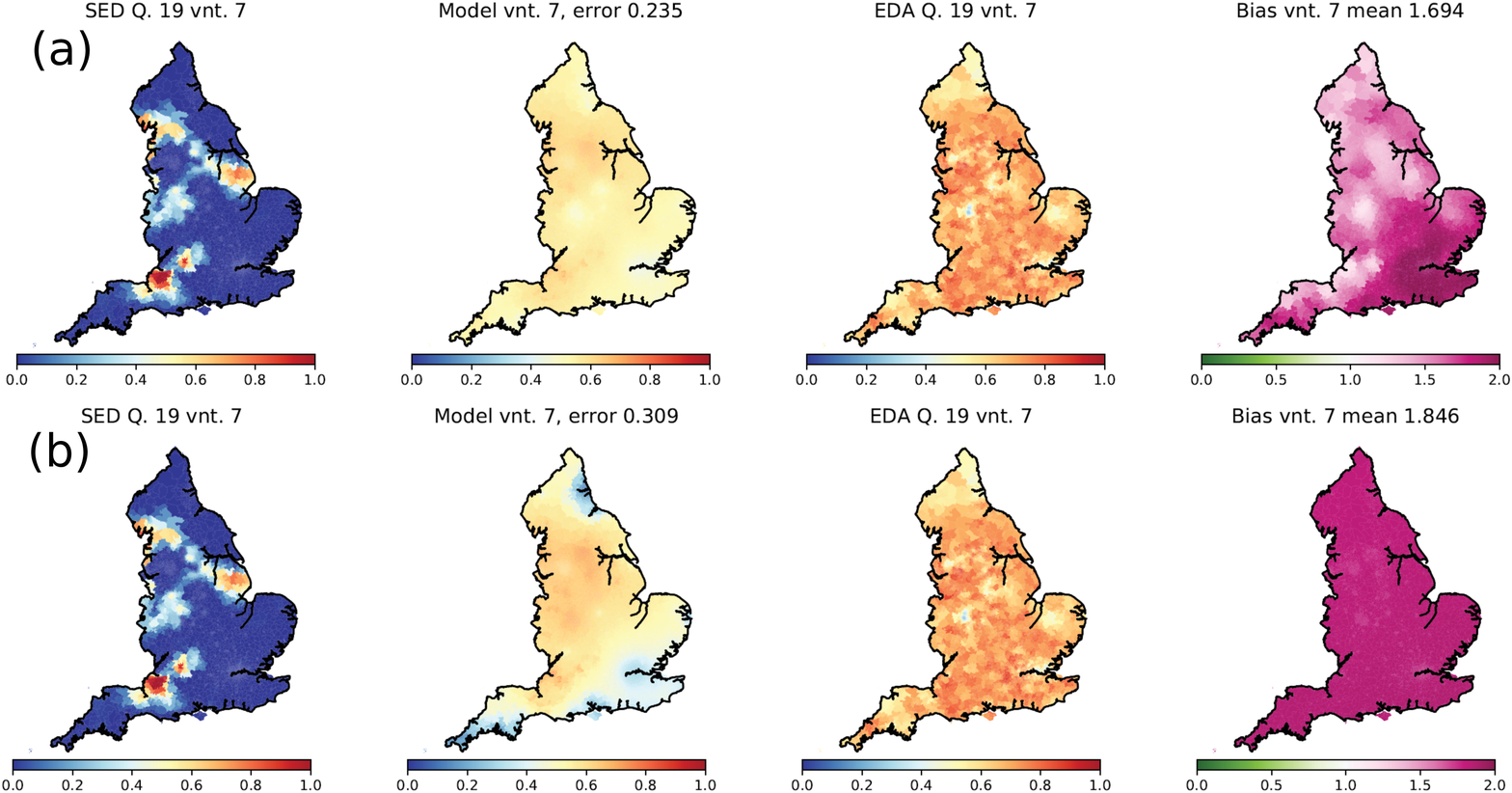}
	\caption{(a) SED (left),  bias model results (middle left) and EDA (middle right) distributions for variant 7, \textipa{[aU]} of variable 19, reflex of ME \textipa{/u:/}. Model parameters $\lambda_M = 0.1, \lambda_R=0.02, \beta=2.0$. Far right map shows calibrated bias field. (b) As for (a), but using $\sigma_s=150$km to yield constant bias field.}
	\label{fig:q19_bias}
\end{figure}

In several cases the no-bias model would fail to predict levelling at all, or would even predict levelling in the opposite direction to that observed. Here the direction of levelling is driven by bias: some variant-specific external or internal factor such as normative bias or markedness must be responsible for the rise in the successful variant. An example of the first type is variable 19 (Figure \ref{fig:q19}): the reflexes of ME \textipa{/u:/} (generally Wells' \textsc{mouth}). Although the no-bias model correctly predicts that small variants (\textipa{[@U]}, \textipa{[E:]}, \textipa{[aI]}) are largely levelled out, since the major variants at the time of the SED (\textipa{[u:]}, \textipa{[EU]}, \textipa{[\ae U]}, \textipa{[a:]} and to a lesser extent \textipa{[aU]}) were found in complex domains with none in the majority, the no-bias model incorrectly predicts a great deal of mixing without much change in relative proportions (i.e. without levelling). A powerful bias term (Figure \ref{fig:bias_hist}) is needed to predict the observed near-complete levelling to \textipa{[aU]} (see Figure \ref{fig:q19_bias}).
In a different linguistic domain, we consider the initial element of the 3sg.m. reflexive pronoun (variable 14). Users of the standard variant, \textit{him(self)}, represented a small minority in the SED and the no-bias model predicts that this variant would disappear; instead, the opposite happened, with substantial levelling towards \textit{him(self)} in the EDA data. In the model, this requires strong bias in favour of \textit{him(self)} (see Figure \ref{fig:bias_hist}).

\subsubsection{Isogloss positioning}

\begin{figure}
	\centering
	\includegraphics[width=\linewidth]{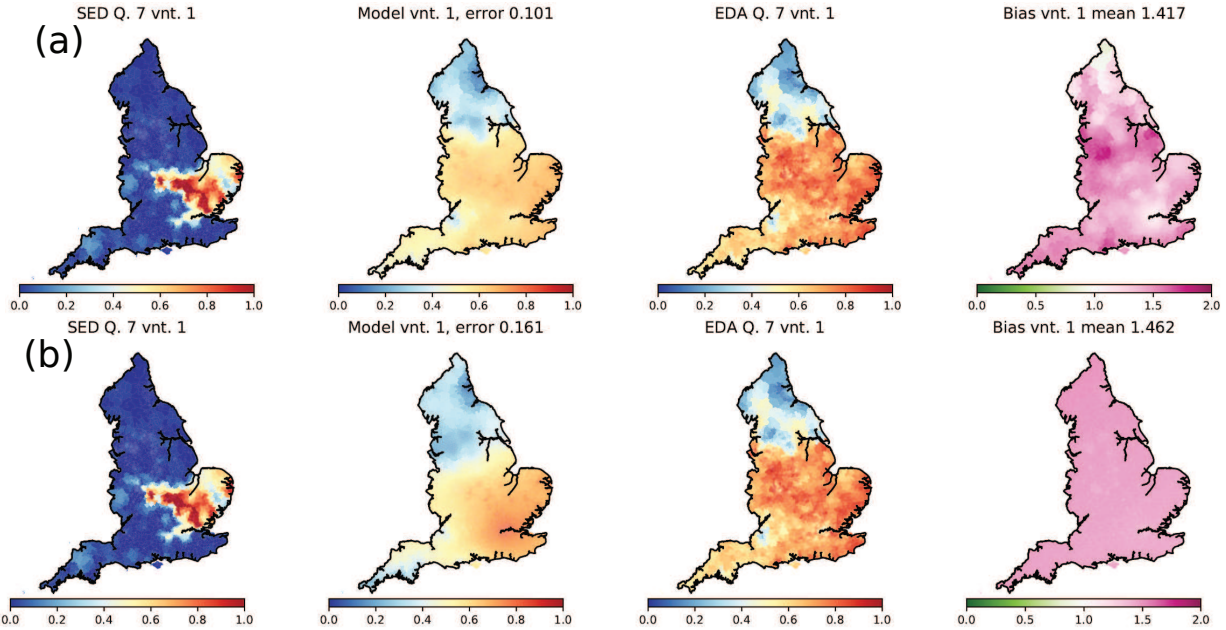}
	\caption{(a) SED (left),  bias model results (middle left) and EDA (middle right) distributions for variant \textipa{/r/} of variable 7, intrusive r, with model parameters $\lambda_M = 0.1, \lambda_R=0.02, \beta=2.0$. Far right map shows calibrated bias field. (b) As for (a), but using $\sigma_s=150$km to yield constant bias field. }
	\label{fig:q7_v1_bias_results}
\end{figure}

In certain cases, although the direction of change is driven by positive bias throughout the spatial domain, the model correctly predicts the new location of an isogloss as a product of coastline and population density. In Figure \ref{fig:q7_v1_bias_results}, the model captures the movement northward of the intrusive r isogloss so that its current position divides the country in a line stretching from the Humber estuary to the Ribble estuary (the result of coastline indentations) with irregularly shaped differentiation within the northern area (the result of population density patterns). Here the spatially varying and spatially constant bias field produce very similar results (Figure \ref{fig:q7_v1_bias_results} (a) and (b)) indicating that isogloss positioning is influenced in a partially predictable way by geography and population distribution, as predicted in refs. \cite{bur17} and \cite{bur18}. The same isogloss appears (spuriously) in the bias free evolution of variable 10 (Figure \ref{fig:q10_no_bias_results}).

\subsubsection{Accelerating spatially driven changes}

\begin{figure}
	\centering
	\includegraphics[width=\linewidth]{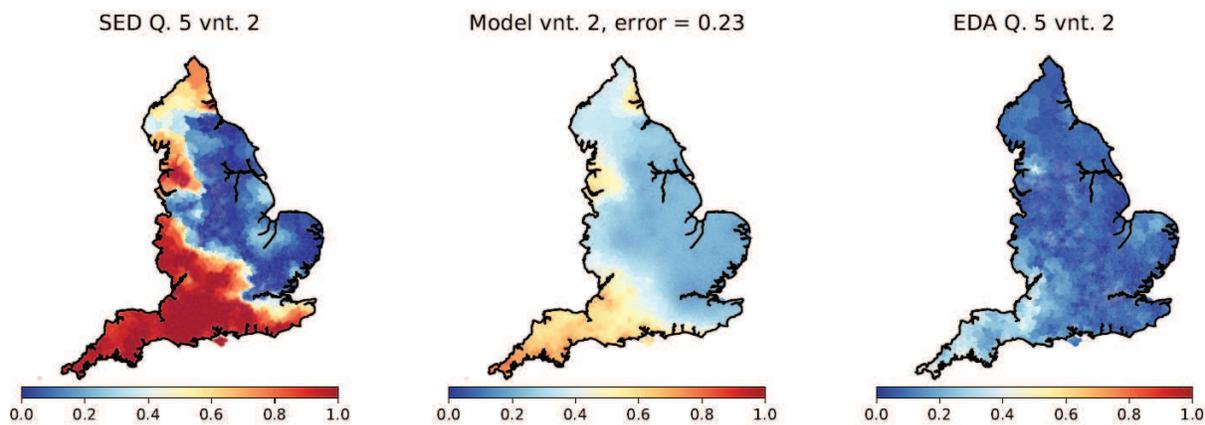}
	\caption{SED (left),  no-bias model results (middle) and EDA (right) distributions for variant \textipa{/r/} of variable 5, rhoticity, with model parameters $\lambda_M = 0.1, \lambda_R=0.02, \beta=2.0$}
	\label{fig:q5_v2_no_bias_results}
\end{figure}

For some variables, a pure spatial model would predict the direction of change correctly, but at a slower pace than observed: here the bias term is needed to increase the rate of change. An example is the changes affecting the circumstances in which the rhotic consonant /\textipa{r}/ is pronounced (variable 5): the no-bias model correctly predicts that the rhotic-non-rhotic isogloss in the south is retreating westwards and that rhoticity is becoming less categorical within the rhotic regions, but greatly underestimates this trend; bias in favour of non-rhoticity is required to explain how advanced the change is (see Figure \ref{fig:q5_v2_no_bias_results}). Another example is the levelling of the 3sg.f. possessive pronoun (variable 15). The no-bias model correctly predicts that the traditional southern and Midlands variant \textit{hern} is levelled in favour of standard \textit{hers} to the extent that no clear isogloss is left, but incorrectly predicts remaining hotspots of traditional usage in high population density areas (Birmingham, Portsmouth/Southampton, Bristol). Bias towards the standard variant accelerates the change in the model, so that these hotspots too are levelled out. A more striking example is variable 8 (the word for \textit{autumn}) where the no bias model predicts that the \textit{autumn-backend} isogloss will move $\approx 100$km north, whereas in reality \textit{backend} has all but disappeared, being used by only $\approx 20\%$ of the population in the far north.

\subsubsection{What does bias represent?}

\begin{figure}
	\centering
	\includegraphics[width=\linewidth]{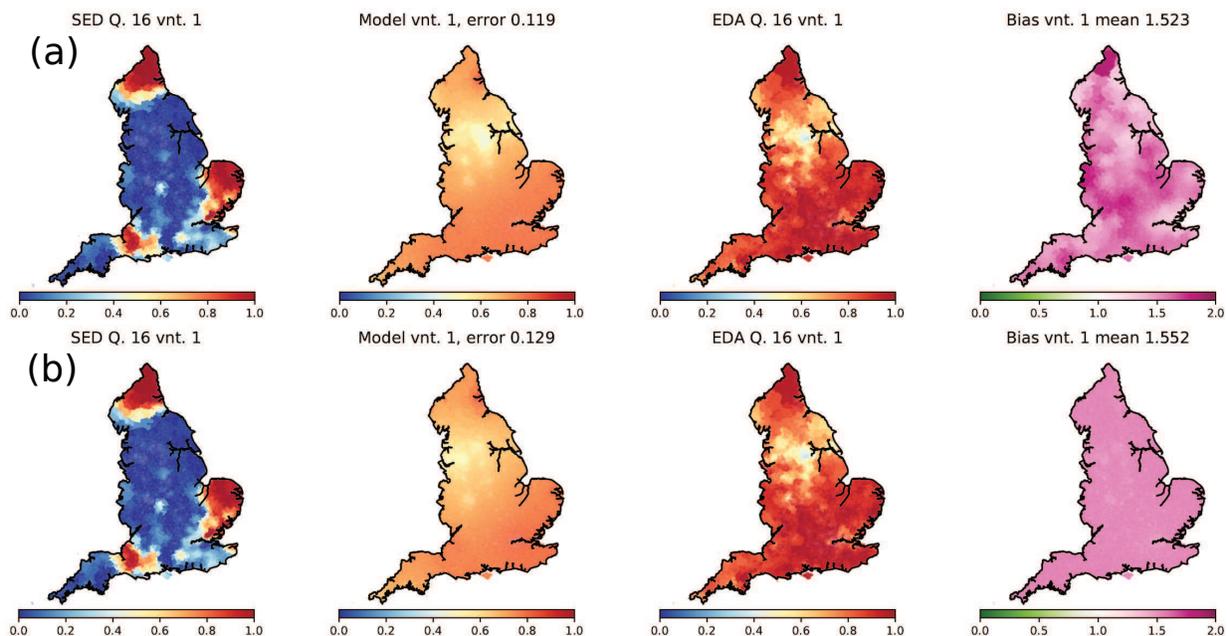}
	\caption{(a) SED (left),  bias model results (middle left) and EDA (middle right) distributions for variant \textipa{[h]} of variable 16, h-dropping, with model parameters $\lambda_M = 0.1, \lambda_R=0.02, \beta=2.0$. Far right map shows calibrated bias field. (b) As for (a), but using $\sigma_s=150$km to yield constant bias field.  }
	\label{fig:q16_v1_bias_results}
\end{figure}

In some cases bias plausibly reflects the existence of a phonetically natural direction of change (the pronunciation of ``l'' in shelf shifting from clear, through dark, to vocalised [\textipa{l}], for example,  but not the reverse). In principle, interaction between different dialect features could even result in such internal bias being spatially variable, although there are no absolutely convincing examples of this in these results.  In rather more cases, however, it seems most likely that bias represents external (non-linguistic) factors and in particular the standard ideology. We can see this wherever bias favours a long-standing standard variant. Examples include the 3sg.m. reflexive pronoun (variable 14), rhoticity (variable 5), and the 3sg.f. possessive pronoun (variable 15). Another example is 
the decrease in ``h dropping'' (variable 16) between the SED and EDA (see Figure \ref{fig:q16_v1_bias_results}); because this is a phonologically unnatural change which undoes a simplification of the pronunciation system, the bias in the model in favour of \textipa{[h]} may reflect the normative pressure of the standard (\textipa{[h]} has gained the status of \textit{standard variant}). These results are inconsistent with an account which centres on mobility and convergence as necessary and sufficient explanatory factors for levelling. Instead, other, variant-specific factors---primarily external factors such as the standard ideology---are required.

A final pattern worth discussing is that of \textit{innovation diffusion}. In four variables, we can plausibly label the changes we see between the SED and EDA as the expansion of a recent innovation: the \textipa{[f]} variant of \textipa{/T/} (variable 6); the \textipa{[P]} variant of coda \textipa{/t/} (variable 22); the \textipa{[\textltilde]} and \textipa{[U]} variants of coda \textipa{/l/} (variable 9); and intrusive r (variable 7). Any account of the dynamics of language change---including non-spatial accounts---must offer an explanation for how innovations take hold. An innovation, by definition, begins as an extreme minority variant, and so we are faced with a challenge to explain why such variants are not immediately levelled in favour of the overwhelming majority. Possible explanations include internal linguistic reasons why particular variants might be favoured (although such explanations must also answer the question of why the change had not already taken place), and social explanations such as age vectors and incrementation or divergence for the purpose of group identity formation \cite{sta16,mich19,kau20}. This is a larger question and not one that this paper seeks to answer; in our models, this too is simply subsumed under the bias term.

\subsection{Time series}

\begin{figure}
	\centering
	\includegraphics[width=\linewidth]{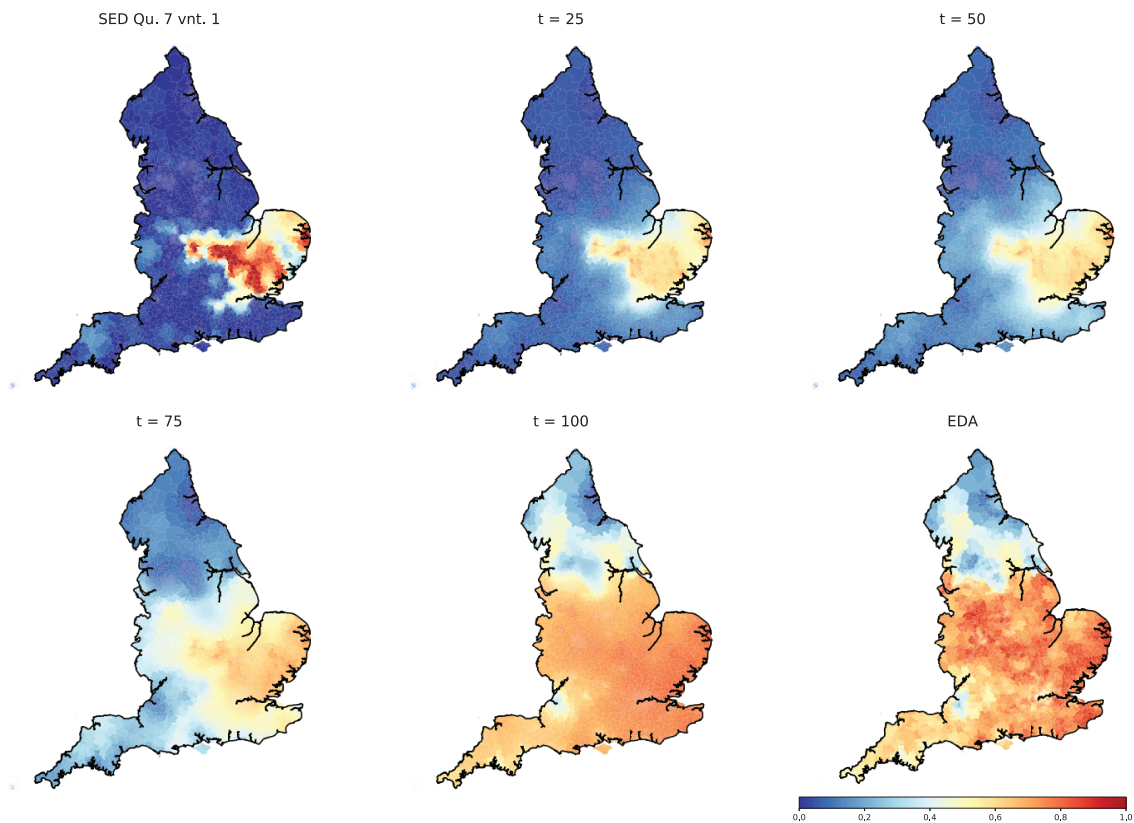}
	\caption{One hundred year time series (in 25 year steps) for variant \textipa{/r/} of variable 7, \textit{intrusive r} in ``thawing'', with model parameters $\lambda_M = 0.1, \lambda_R=0.02, \beta=2.0$.}
	\label{fig:time_series}
\end{figure}

\begin{figure}
	\centering
	\includegraphics[width=\linewidth]{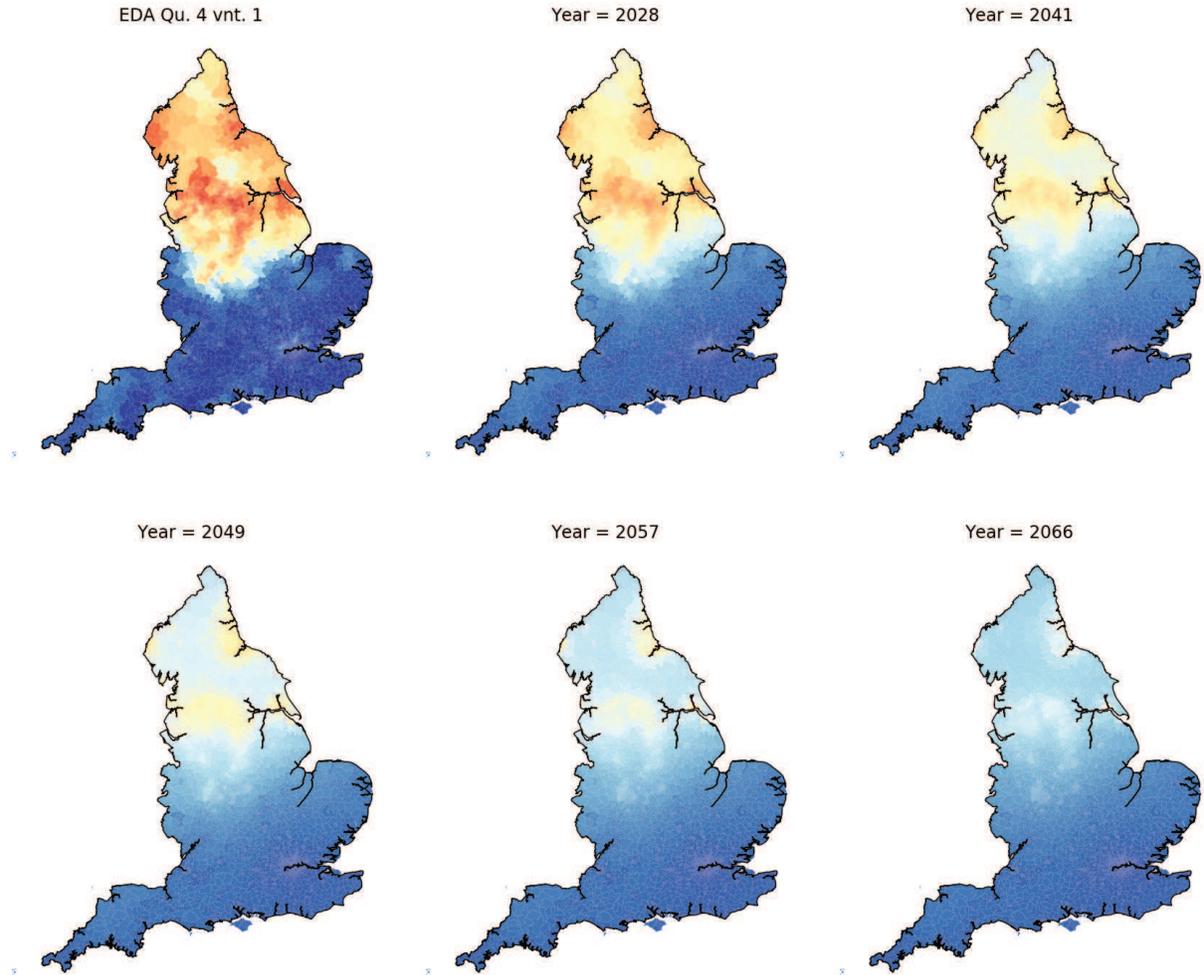}
	\caption{Predicted future of \textsc{foot-strut}. Maps show predicted future of variant 1 [\textipa{U}] of variable 4 (Table \ref{tab:vars}); the vowel in ``butter''. Bias field calibrated to SED/EDA using $\sigma_s=150$km, and model initialized with EDA. Mean calibrated bias parameters $\bv{h}=(0.71,1.29)$, and model parameters $\lambda_M = 0.1, \lambda_R=0.02, \beta=2.0$. Start year = 2016 (date of EDA survey).}
	\label{fig:future}
\end{figure}

So far we have presented only the initial and final states generated by our model, but we can also provide a complete reconstruction of the history of each variant over the 20th Century. For example, Figure \ref{fig:time_series} shows our predictions for the evolution of \textit{intrusive r} in ``thawing''. In principle such time series can be compared to mid-century data to validate our reconstruction of the historical changes which have taken place. Such reconstructions may have utility in historical linguistics, which aims to understand the sequences of linguistic changes which have led to the current state of the world's languages. If we were to make an initial guess at a historical language state, a model similar to ours, adapted to allow for sequential changes in variants, could be calibrated to later states to provide a predicted history. This could then be compared to predictions made by more traditional linguistic methods. In principle it would be possible to impose additional constraints on the model, set by further data, at intervening points in its evolution. Where changes are still in progress, the model might also be used to predict the future. As an example, we have calibrated a constant bias field for variable 4 (\textsc{foot-strut}), giving model error 0.106. We then used this field to evolve the model into the future, using the EDA as initial condition. The results are shown in Figure \ref{fig:future}, up to the year 2066. Constant bias was used to reduce the chance of over fitting.

\section{Conclusion}

We have developed an explicit spatial model of language evolution, accounting for mundane daily mobility, realistic migration patterns, spatial distribution of population, and social conformity. One motivation is that many scholars examining diffusion and levelling in English English over the last century have argued for the primacy of movement (both ``routine'' mobility and migration) in driving change (cf. section \ref{sec:linguistic background}). This---tacitly or explicitly---implies that ideological and normative factors and internal linguistic pressures play only secondary roles: that widespread levelling would have taken place even without the strong standard ideology that characterises English language attitudes, and that this ideological context served only to hasten or to determine the precise direction of processes which were the product of more fundamental social-technological changes. In scientific terms, we can think of the extreme form of this view,  where all change is driven by purely spatial processes (diffusion, long range mixing, interface motion), as a \textit{null model} - the variant symmetric form of our model. We have shown that although the evolution of some variants can be explained in this way, the majority of historical changes could not have occurred without certain variants gaining special status. In some cases this bias merely accelerates or further drives changes which would have partially occurred by spatial effects alone, requiring a relatively low level of asymmetry. In other cases, strong biases are required either to drive minority variants such as \textit{him(self)} to national dominance, or to preserve distinctive local forms such as \textit{spelk}. In most cases the origins of these biases appear more likely to be external social factors such as normative standards, rather than internal linguistic factors. We have also seen that, even when bias is needed to explain observations, spatial effects often play an important role. Our conclusion is therefore that spatial process are important, but insufficient to explain the majority of the changes we have seen. A complete catalogue of results for all variants, including bias fields may be found in the supplementary material \cite{bur21}.

Beyond assessing the importance of spatial processes to language change, this modelling exercise represents a step towards the construction of minimalist spatial models which can be used to reconstruct more general historical linguistic changes, and potentially predict future change. In providing a complete hypothetical time series for a long period of time, we are offering explicit predictions which can be compared to data collected in England at any point through the century. Since the bias field is not an observable quantity, it is only these intermediate states that are directly falsifiable. We hope that such comparisons may be used to test,  improve or refute our model in future. 

\subsection*{Code and data availability}

No new data were created or analysed in this study. Burridge is happy to be contacted with reasonable requests for additional maps, or source code.

\subsection*{Acknowledgements}

The authors thank the referees for their diligent and insightful reviews, which raised a number of important questions. They are also grateful for the support of a Royal Society APEX award APX\textbackslash R1\textbackslash 180117 (funded by the Leverhulme Trust). 
 
\bibliographystyle{unsrt}
\bibliography{SED_EDA_refs}

\end{document}